\documentclass[twocol]{ametsocV6.1}

\title{An analytical model for tropical cyclone outer-size expansion on the $f$-plane \color{red}NOT PUBLISHED. In peer review}

\authors{Danyang Wang \aff{a},\correspondingauthor{Danyang Wang, wang5571@purdue.edu}
Daniel R. Chavas\aff{a} 
}

\affiliation{\aff{a}{Department of Earth, Atmospheric, and Planetary Sciences, Purdue University, West Lafayette, Indiana}}

\abstract{
Tropical cyclones are known to expand to an equilibrium size on the $f$-plane, but the expansion process is not understood. In this study, an analytical model for tropical cyclone outer-size expansion on the $f$-plane is proposed.  
Conceptually, the storm expands because the imbalance between latent heating and radiative cooling drives a lateral inflow that imports absolute vorticity. 
Volume-integrated latent heating increases more slowly with size than radiative cooling, and hence the storm expands towards an equilibrium size.
The predicted expansion rate is given by the ratio of the difference in size from its equilibrium value ($r_{t,eq}$) to an environmentally-determined time scale $\tau_{rt}$  of $10\sim15$ days. The model is fully predictive if given a constant $r_{t,eq}$, which can also be estimated environmentally. 
The model successfully captures the first-order size evolution across a range of numerical simulation experiments in which the potential intensity and $f$ are varied. 
The model predictions of the dependencies of lateral inflow velocity and expansion rate on latent heating rate also compare well with numerical simulations. 
This model provides a useful foundation for understanding storm size dynamics in nature.
}

\begin{document}

\maketitle

%
%
%
%
%
%

%








\section{Introduction}
Tropical cyclone (TC) outer size is known to expand with time towards an equilibrium size in idealized simulations on the $f$-plane (Chan and Chan 2014, 2015; Chavas and Emanuel 2014; Martinez et al. 2020\nocite{Chan_Chan_2014, Chan_Chan_2015b,Chavas_Emanuel_2014,Martinez_Nam_Bell_2020}). Expansion with time is also seen in reanalysis or simulations on spherical geometry (Schenkel et al. 2018, 2023\nocite{Schenkel_etal_2018,Schenkel_et_al_2023}). Reanalysis data shows the median expansion rate of TC outer radius (of 8 m/s near-surface wind) is tens of kilometers per day, with extreme cases being hundreds of kilometers per day (Schenkel et al. 2023).  Although TC intensity and intensification have been understood with the help of some relatively well established analytical theory (Emanuel 1986, 2012; Emanuel and Rotunno 2011; Wang et al. 2021a,b\nocite{Emanuel_1986,Emanuel_2012b,Wang_et_al_2021,Wang_Li_Xu_2021}), a conceptual understanding of tropical cyclone size and size expansion is not as complete. 
Although theoretical models links inner size (radius of maximum wind) to outer size (Emanuel and Rotunno 2011; Chavas and Lin 2016\nocite{Emanuel_Rotunno_2011,Chavas_Lin_2016}), the mechanism of the changes of inner and outer sizes are not the same (Weatherford and Gray 1988; Chavas and Knaff 2023\nocite{Weatherford_Gray_1988,Chavas_Knaff_2022}); the present study will focus on TC outer-size expansion mechanism.

Recently, Wang et al. (2022\nocite{Wang_et_al_2022}) proposed a model for tropical cyclone potential size (TC PS) on the $f$-plane that explains equilibrium TC size and is solely dependent on environmental parameters. The model yields a new scaling that is similar to the length-scale $V_p/f$, where $V_p$ is the potential intensity and $f$ the Coriolis parameter, proposed in prior work \citep{Chavas_Emanuel_2014}. The TC PS model combines the Carnot cycle model for the energetics of a TC (Emanuel 1988, 1991\nocite{Emanuel_1988,Emanuel_1991}) and a model for the complete low-level TC wind field (Chavas et al. 2015\nocite{Chavas_Lin_Emanuel_2015}) 
to solve for an equilibrium size based on the most efficient thermodynamic cycle.
However, such a method does not provide a description of how other parts of the TC are working, without which the potential size may not be achieved at all. It is also unsatisfying that the thermodynamic cycle is formulated in steady state so that it does not mechanistically explain how and why a TC expands towards equilibrium. Although the model suggests that an energy surplus exists when a TC is smaller than its potential size, it cannot explain how this energy surplus might drive expansion.

Previous studies on TC size expansion consistently note the importance of low-level inflow for bringing environmental absolute angular momentum inwards to drive expansion (Hill and Lackman 2009\nocite{Hill_Lackmann_2009}; Bui et al. 2009\nocite{Bui_Smith_Montgomery_Peng_2009}; Wang 2009\nocite{Wang_2009}; Chan and Chan 2014, 2015, 2018\nocite{Chan_Chan_2014,Chan_Chan_2015b,Chan_Chan_2018}; Martinez et al. 2020\nocite{Martinez_Nam_Bell_2020}; Wang and Toumi 2022\nocite{Wang_Toumi_2022}), which is a direct reflection of the spinup of the outer-core wind field. TC size expansion rate has been further found to depend on initial vortex size (Xu and Wang 2010\nocite{Xu_Wang_2010}; Chan and Chan 2014; Martinez et al. 2020\nocite{Chan_Chan_2014,Martinez_Nam_Bell_2020}), rainband activity (Hill and Lackman 2009; Wang 2009; Fudeyasu et al. 2010; Martinez et al. 2020\nocite{Hill_Lackmann_2009,Wang_2009,Fudeyasu_2010,Martinez_Nam_Bell_2020}), as well as cloud radiative forcing (Bu et al. 2014, 2017\nocite{Bu_Fovell_Corbosiero_2014,Bu_Fovell_Corbosiero_2017}). 
Simulations have also shown that TC size is able to continue expanding substantially long after intensity becomes quasi-steady (Hill and Lackman 2009; Chan and Chan 2014, 2015; Martinez et al. 2020\nocite{Hill_Lackmann_2009,Chan_Chan_2014,Chan_Chan_2015b,Martinez_Nam_Bell_2020}).
However,  a simple universal understanding of why a TC should expand, how fast, why size should approach an equilibrium, and how this behavior depends on environmental parameters is still lacking. This is partly because the lateral inflow or import of absolute angular momentum has yet to be fully and quantitatively linked to environmental parameters and internal processes. 
Such a quantitative link, either direct or indirect, is necessary for a predictive model for size. Indeed, if given an inflow velocity, then size expansion may be predicted as shown in \cite{Wang_Toumi_2022}.
However, inflow velocity varies significantly with height, from larger values within the boundary layer to near zero at some height above the top of boundary layer. Thus, one needs to consider the integrated inflow mass flux instead of picking one single height. 

In this study, we propose a model for size expansion towards equilibrium on the $f$-plane, in terms of the outer radius of a certain tangential wind speed at the top of boundary layer.
\footnote{For TC size, one may consider a single outer wind radius because the wind field structure is fully specified from a single input size (Chavas et al. 2015).}
We seek a model that is
\begin{itemize}
    \item predictive and analytic;
    \item yields a characteristic expansion rate from environmental/external parameters;
    \item explains the physical process that drives TC expansion, and why this expansion vanishes such that there exists an upper bound of size.
\end{itemize}
We test these model outcomes via comparison of model predictions with sets of numerical simulation experiments. 


Our model for TC size expansion  is presented in Sec. 2. 
Basic predictions of the model and its comparison to numerical simulations are provided in Sec. 3. 
Further physical interpretations of the model are provided in Sec. 4.
A summary of key conclusions and discussion is given in Sec. 5.

\section{Theory: An analytical outer-size-expansion model on the $f$-plane}
\subsection{Basic model structure}

\begin{figure}[t]
\centering
 \noindent\includegraphics[width=0.8\textwidth,angle=0]{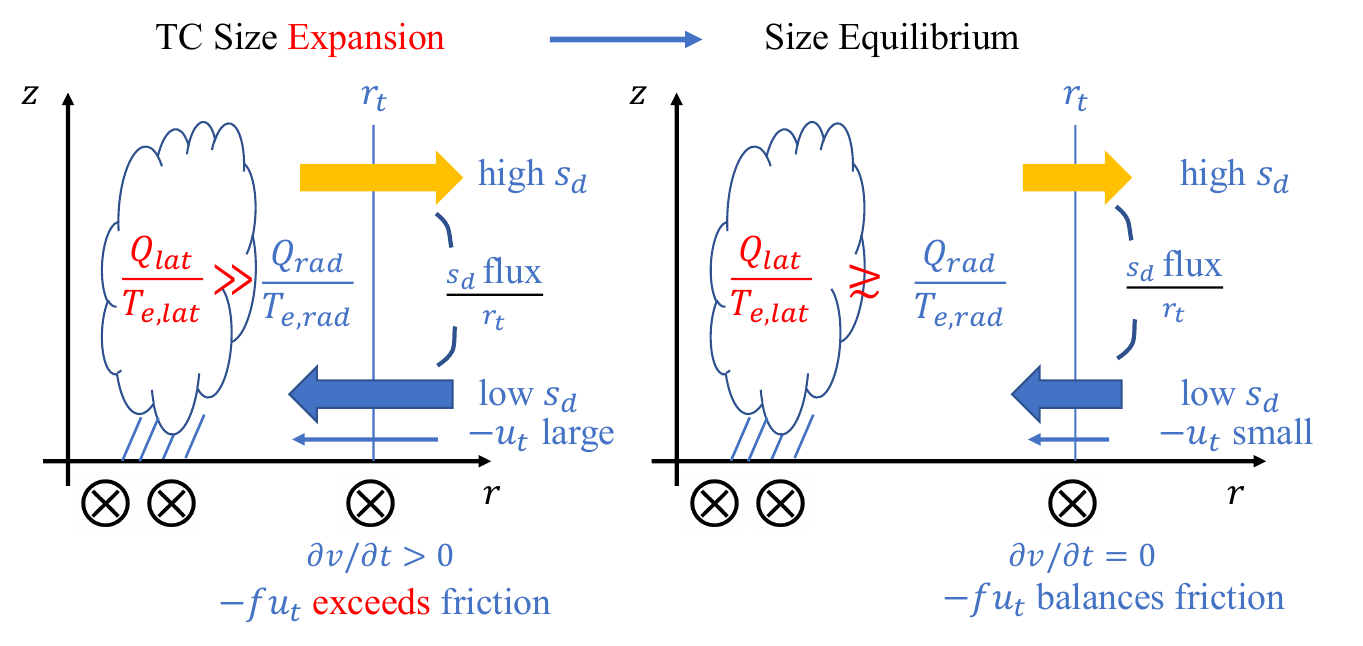}\\
 \caption{A schematic plot of the expansion model presented in Sec. 2. See text for details.
 }\label{fig:schematic}
\end{figure}

Below we present a theory for TC expansion towards equilibrium that is summarized conceptually as follows:
1) In radiative-convective equilibrium (RCE) without a TC, net condensational heating equals net radiative cooling; 
2) When a TC forms, the TC volume is shifted substantially out of RCE, such that condensational heating substantially exceeds radiative cooling (consistent with enhanced surface fluxes); 
3) The TC expands in response as a result of strong low-level inflow as part of the overturning circulation that exports excess heat. As it expands, area-integrated radiative cooling increases faster than net condensational heating until low-level inflow is weak enough so that surface friction prevents any further expansion of wind field. The storm has reached its equilibrium size.
A schematic plot is shown in Fig. \ref{fig:schematic}.

\clearpage

\clearpage

We define $r_t$ as the radius of a fixed tangential velocity $v_t$ (for example $r_8$ the radius of $v_t=8$ m/s  tangential wind) at the top of boundary layer in the TC outer-core region.
As TC size expansion is basically low-level spin-up of TC outer core, the expansion rate of $r_t$ can be given by:
\begin{equation}\label{eq:kinematic}
    \frac{dr_t}{dt}=\frac{\partial v}{\partial t}/(-\frac{\partial v}{\partial r}), \quad r=r_t
\end{equation}
which is obtained by taking $dv/dt=0$ with $v=v(r(t),t)$, with $v$ the tangential wind, $r$ the radius and $t$ the time\footnote{Technically, $r_t$ is understood as $r_t=r(v_t,t,\vartheta)$, where $\vartheta$ represents a series of environmental parameters, and $v_t$ is a time-independent tangential velocity. Since the main focus for expansion rate is with a fixed $v_t$ in a given environment (fixed $\vartheta$), we write $dr_t/dt$ instead of $\partial r_t/\partial t$.}.
This relation is also presented in Tsuji et al. (2016\nocite{Tsuji_etal_2016}).

The local spinup tendency for an axisymmetric TC on the $f$-plane is given by
\begin{equation}
    \frac{\partial v}{\partial t}=-u(f+\zeta)-w\frac{\partial v}{\partial z}+F, \quad r=r_t
\end{equation}
where $\zeta=\frac{\partial v}{\partial r}+\frac{v}{r}$ is the relative vorticity, $u$ the radial velocity, $w$ the vertical velocity and $F\approx\frac{1}{\rho_d}\frac{\partial\tau_v}{\partial z}$ the turbulence frictional force with $\rho_d$ the dry air density and $\tau_v$ the turbulence stress in azimuthal direction. Eq. (2) is simplified by choosing $r_t$ sufficiently far from the center so that $\zeta$ can be neglected compared to $f$. 
Further integrating Eq. (2) from the surface to $h_w$, some height in the lower troposphere below which the inward mass flux constitutes the majority of total lateral inward mass flux, and neglecting vertical advection, gives
\begin{equation}\label{eq:vten2}
    \frac{\partial v}{\partial t}\approx -fu_t-C_d(\mu v_t)^2/h_w
\end{equation}
where the aerodynamic formula for surface stress $\tau_{v,sfc}=\rho_d C_d|\boldsymbol{V}_{10}|v_{10}$ is applied with $C_d$ the surface exchange coefficient for momentum, $\boldsymbol{V}_{10}$ the 10-m surface horizontal velocity, $v_{10}$ the 10-m tangential velocity, $u_t$ the vertical mean radial velocity, $v_t$ the corresponding tangential veclocity at the top of boundary layer at $r_t$, and $\mu$ is a surface wind reduction factor (i.e. basically the ratio of $v_{10}$ to $v_t$). 
In Eq. (\ref{eq:vten2}), the second term on the RHS can be taken as a constant, a key conceptual benefit  since we are following the radius of a fixed wind speed. 
Substituting Eq. (\ref{eq:vten2}) into Eq. (\ref{eq:kinematic}) gives
\begin{equation}\label{eq:drtdt_0}
    \frac{dr_t}{dt}=\frac{1}{(-\frac{\partial v}{\partial r})}\left[-fu_t-C_d(\mu v_t)^2/h_w\right]
\end{equation}
There are two quantities, $\partial v/\partial r$ and $u_t$, that are not specified and must be linked to internal processes or environmental parameters.   
The slope of wind profile $\frac{\partial v}{\partial r}$ can be obtained from the wind profile solution for the outer wind field from Emanuel (2004, here after E04\nocite{Emanuel_2004}) model (see Appendix B), which depends only on environmental parameters. The E04 model does not have an analytical solution, but we will provide an analytic approximation for $\frac{\partial v}{\partial r}$ based on this model in Sec. 2b. An expression for $u_t$ is derived next.

We propose $u_t$ to be driven principally by the energetics of the TC, i.e., latent heating and radiative cooling, which can be described by a dry entropy budget. Here dry entropy is loosely defined by $s_d=c_p\ln\frac{\theta}{T_{trip}}$, with $c_p$ the specific heat of dry air at constant pressure, $\theta$ the potential temperature and $T_{trip}$ the tripple-point temperature; $s_d$ thus defined is a close approximation of true dry entropy, and is more convenient for budget analysis in numerical simulations.
The budget of dry entropy $s_d$ within the TC volume from the center to $r_t$ is written as:
\begin{equation}\label{eq:sdbud}
\frac{\partial \mathcal{S}}{\partial t}=\frac{\mathcal{Q}_{lat}}{T_{e,lat}}-\frac{\mathcal{Q}_{rad}}{T_{e,rad}}+\mathcal{\Dot{S}}_{res}+\mathcal{F}_r+\mathcal{F}_u
\end{equation}
where $\mathcal{S}$ is mass integrated dry entropy within the volume; $\mathcal{Q}_{lat}$ and $\mathcal{Q}_{rad}$ are the net condensational heating (latent heat) and the total radiative cooling (defined positive), respectively, with $T_{e,lat}$ and $T_{e,rad}$ their respective effective temperatures; $\mathcal{\Dot{S}}_{res}$ represents other sources of dry entropy, such as surface sensible heating, diffusion of sensible heat and dissipative heating; and $\mathcal{F}_r$ and $\mathcal{F}_u$ are the fluxes of dry entropy into the volume from the lateral (at $r_t$) and vertical directions (at the upper extent of the volume), respectively\footnote{A dry static energy budget is also viable, and the effective temperatures will not appear so that sensible and latent heat need not be separated. However, we use dry entropy budget because it is more tractable for comparison with numerical simulations.}. The first two terms on the RHS are dry entropy sources due to latent heating and radiative cooling, respectively.
See Appendix D for detailed expressions.
To achieve a simple expression for size expansion, we neglect $\frac{\partial \mathcal{S}}{\partial t}$, $\mathcal{\Dot{S}}_{res}$, $\mathcal{F}_u$. 
This assumes the dominant terms are sources/sinks from latent heating (source), radiative cooling (sink) and net lateral transport into the TC from the environment (supported by simulations in Appendix D). 
Doing so yields the balance equation
\begin{equation}\label{eq:sdbud2}
    \frac{\mathcal{Q}_{lat}}{T_{e,lat}}-\frac{\mathcal{Q}_{rad}}{T_{e,rad}}\approx-\Delta s_d 2\pi r_t \rho_i u_th_w
\end{equation}
where we have rewritten the lateral flux term in terms of a bulk free-tropospheric dry static stability given by
\begin{equation}\label{eq:delta_sd}
    \Delta s_d=\mathcal{F}_r/(2\pi r_t\rho_i u_th_w)
\end{equation}
.  
Though $\Delta s_d$ must also depend on the vertical profile of lateral flow (for which we lack a clear constraint), the physical meaning of $\Delta s_d$ can be better understood in the ideal case where inflow is confined to near the surface and outflow is confined to near the tropopause level: in this case $\Delta s_d$ represents the difference of $s_d$ between the surface and tropopause. Appendix D shows that this is a reasonable assumption for TCs; discussion of the meaning of $\Delta s_d$ in general is also provided in Appendix D. 
The parameter $\rho_i$ is an effective inflow air density corresponding to $u_t$ so that $2\pi r_t\rho_iu_th_w$ is the lateral mass flux at $r_t$ below $h_w$. 
A reference of $\Delta s_d$ is the difference between moist entropy and dry entropy near sea surface, which is equivalent to the difference of $s_d$ between tropopause and surface. 
This assumes the eyewall to be in slantwise neutrality, which applies to the later stage of TC intensification and peak intensity (Bryan and Rotunno 2009; Peng et al. 2018; Wang et al. 2021b), which is the principal period for size expansion to occur (e.g. Martinez et al. 2020). 
For near-surface air with water vapor mixing ratio $q_v=0.018$ kg/kg, temperature $T=300$ K, and relative humidity 80\% (tropical value, see Dunion 2011\nocite{Dunion2011}), this gives a reference $\Delta s_d$ of $L_vq_v/T\approx150$ J/K/kg, where $L_v=2.501\times10^{6}$ J/kg is the latent heating of vaporization. Thus, $\Delta s_d$ can be taken as primarily determined by sea surface temperature.

Latent heating is assumed to be principally produced in the eyewall (see Appendix D), which is largely driven by boundary layer frictional convergence as found in both observations of vertical velocities (Stern et al. 2016\nocite{Stern_etal_2016}) and implicit in the slantwise neutrality assumption of potential intensity theory (Emanuel 1986, 1995; Khairoutdinov and Emanuel 2013\nocite{Emanuel_1986,Emanuel_1995a,Khairoutdinov_Emanuel_2013}).  Hence, $\mathcal{Q}_{lat}$ may be written as
\begin{equation}\label{eq:Qlat}
    \mathcal{Q}_{lat}=\frac{\epsilon_{p,ew}}{\alpha_p}\mathcal{Q}_{c,ew}  \approx \frac{\epsilon_{p,ew}}{\alpha_p} L_v q_{vb}{M}_{ew}
\end{equation}
where $q_{vb}$ is boundary layer water vapor mixing ratio just outside of the eyewall corresponding to $M_{ew}$; $\epsilon_{p,ew}$ is the precipitation efficiency in the eyewall region, defined as the ratio of condensation to the mass of water vapor imported upwards into the eyewall (see Appendix D for practical diagnosis); $\mathcal{Q}_{c,ew}$ is the total condensation rate in the eyewall; $\alpha_p
$ the ratio of net latent heating in the eyewall region to that within $r_t$; and $M_{ew}$ the eyewall updraft mass flux. 
Given that the eyewall updraft is driven by boundary layer frictional convergence, $M_{ew}$ is also equal to the friction-induced inflow mass flux into the eyewall.
Thus, $M_{ew}/\rho_w$ (which will appear shortly) should be strongly controlled by inner-core size and TC intensity, where 
$\rho_w$ is the effective dry air density for boundary layer inflow under the eyewall (close to $\rho_i$, see Appendix C for calculation).
Here $\rho_w$ becomes implicit, as in the boundary layer momentum equations only the gradient wind matters and air density will not explicitly appear (Kuo 1982; Kepert 2001). 

Radiative cooling $\mathcal{Q}_{rad}$ may be written as (Chavas and Emanuel 2014\nocite{Chavas_Emanuel_2014})
\begin{equation}\label{eq:Qrad}
    \mathcal{Q}_{rad}=\pi r_t^2 c_p \frac{\Delta p}{g}Q_{cool}
\end{equation}
where
\begin{equation*}
    \Delta p=\frac{p_0}{R/c_p+1}\left[\left(\frac{p_s}{p_0}\right)^{1+R/c_p}-\left(\frac{p_t}{p_0}\right)^{1+R/c_p}\right]
\end{equation*}
In the above, $Q_{cool}$ is a constant radiative cooling rate for potential temperature, $\Delta p/g$ the effective mass obtained by vertical integration over a pressure layer, with $g$ the gravitational acceleration, $p_0=1000$ hPa the reference pressure, $R$ the gas constant of dry air, and $c_p$ the heat content of dry air at constant pressure. 
Taking surface pressure $p_s=1000$ hPa and tropopause pressure $p_t=100$ hPa with $Q_{cool}=1$ K/day yields a value of 88 W/m$^2$, close to the 100 W/m$^2$ value in tropics (Pauluis 2000\nocite{Pauluis_etal_2000}). 

An expression for inflow velocity is obtained by first rearranging Eq. (\ref{eq:sdbud2}): 
\begin{subequations}\label{eq:ut_energetics}
\begin{equation}
    -u_t=\frac{1}{2\pi r_t h_w\rho_i}(\frac{\mathcal{Q}_{lat}}{T_{e,lat}\Delta s_d}-\frac{\mathcal{Q}_{rad}}{T_{e,rad}\Delta s_d})
\end{equation}
then substituting for $\mathcal{Q}_{lat}$ using Eq. (\ref{eq:Qlat}) and $\mathcal{Q}_{rad}$ using Eq. (\ref{eq:Qrad}) to yield
\begin{equation}
    -u_t=\frac{1}{2\pi h_w}\frac{\epsilon_{p,ew}}{\alpha_p}\frac{L_vq_{vb}}{T_{e,lat}\Delta s_d}(\frac{M_{ew}}{\rho_w})\frac{1}{r_t}-\frac{1}{2h_w}c_p\frac{\Delta p}{\rho_i g}\frac{Q_{cool}}{T_{e,rad}\Delta s_d}r_t,
\end{equation}
where we take $\rho_i\approx\rho_w$. We may write this more compactly as 
\begin{equation}
    -u_t=\frac{1}{h_w}A(\frac{M_{ew}}{\rho_w})\frac{1}{r_t}-\frac{1}{h_w}B r_t
\end{equation}
\end{subequations}
where we define two thermodynamic parameters
\begin{subequations}\label{eq:AB}
\begin{equation}
    A=\frac{1}{2\pi}\frac{\epsilon_{p,ew}}{\alpha_p}\frac{L_vq_{vb}}{T_{e,lat}\Delta s_d}
\end{equation}
and
\begin{equation}
    B=\frac{1}{2}c_p\frac{\Delta p}{\rho_ig}\frac{Q_{cool}}{T_{e,rad}\Delta s_d}
\end{equation}
\end{subequations}
. Parameter $A$ is non-dimensional and is related to the latent heating that drives expansion, while $B$ is a velocity and is related to radiative cooling that suppresses expansion.

The size-expansion model is obtained by substituting Eq. (\ref{eq:ut_energetics}c) into Eq. (\ref{eq:drtdt_0}) to yield
\begin{equation}\label{eq:exmodel_general}
    \frac{dr_t}{dt}=\frac{r_{t,eq}-r_t}{\tau_{rt}}
\end{equation}
Here $r_{t,eq}$ is the equilibrium size when $dr_t/dt=0$ is achieved (Eq. \ref{eq:exmodel_general}) and $\tau_{rt}$ the time scale for expansion. 
Eq. (\ref{eq:exmodel_general}) states that expansion rate is given by the difference in size from equilibrium divided by a time scale $\tau_{rt}$. 

Quantity $r_{t,eq}$ in Eq. (\ref{eq:exmodel_general}) is given by
\begin{subequations}\label{eq:rteq}
\begin{equation}
    r_{t,eq}=[fA(\frac{M_{ew}}{\rho_w r_t})-C_d(\mu v_t)^2]/(fB)
\end{equation}
which can be expressed explicitly by $\mathcal{Q}_{lat}$ and $Q_{cool}$ as (using Eqs. \ref{eq:Qlat}, \ref{eq:AB})
\begin{equation}
\begin{split}
    &\quad r_{t,eq}\\
    &=[f\frac{1}{2\pi r_t\rho_i}\frac{\mathcal{Q}_{lat}}{T_{e,lat}\Delta s_d}-C_d(\mu v_t)^2]/(\frac{1}{2}fc_p\frac{\Delta p}{\rho_i g}\frac{Q_{cool}}{T_{e,rad}\Delta s_d})
\end{split}
\end{equation}
\end{subequations}
Eq. (\ref{eq:rteq}a,b) indicates that $r_{t,eq}$ may vary with $r_t$ (and thus time), but here we will take it to be a constant in order to seek an analytical solution of Eq. (\ref{eq:exmodel_general}); this assumption is later tested in Sec. 3b. 
A useful form of $r_{t,eq}$ is obtained by writing Eq. (\ref{eq:rteq}a) at equilibrium ($r_t=r_{t,eq}$): 
\begin{equation}\label{eq:rteq2}
    r_{t,eq}=[fA (\frac{M_{ew}}{\rho_w})_{eq}\frac{1}{r_{t,eq}}-C_d(\mu v_t)^2]/(B f)
\end{equation}
where subscript "eq" means equilibrium. 
Before solving for $r_{t,eq}$, we first define the equilibrium radius of zero net source of dry entropy, $r_{RCE,eq}$, inside of which the system is in radiative-convective equilibrium (RCE), by taking the LHS of Eq. (\ref{eq:sdbud2}) to be zero (using Eqs. \ref{eq:Qlat} and \ref{eq:Qrad}) and solving for $r_t$ at equilibrium:
\begin{equation}\label{eq:r0_eq}
    r_{RCE,eq}=\sqrt{\frac{A}{B}(\frac{M_{ew}}{\rho_w})_{eq}}
\end{equation}
Thus $r_{RCE,eq}$ scales with $\sqrt{(M_{ew}/\rho_w)_{eq}}$ (this relationship will be revisited later).
Note $r_{RCE,eq}$ cannot be obtained by directly taking $v_t=0$ in Eq. (\ref{eq:rteq2}) because $v_t=0$ implies $u_t=0$ in equilibrium (Eq. \ref{eq:drtdt_0}) but Eq. (\ref{eq:delta_sd}) does not allow $u_t=0$. 
Substituting Eq. (\ref{eq:r0_eq}) into Eq. (\ref{eq:rteq2}) and solving for $r_{t,eq}$ we have 
\begin{subequations}\label{eq:rteq_solved}
\begin{equation}
    r_{t,eq}=\frac{-C_d(\mu v_t)^2\frac{1}{B f}+\sqrt{C_d^2(\mu v_t)^4\frac{1}{B^2f^2}+4r_{RCE,eq}^2}}{2},
\end{equation}
which can be written compactly as 
\begin{equation}
    r_{t,eq}=\frac{-\xi v_t^2+\sqrt{\xi^2v_t^4+4r_{RCE,eq}^2}}{2}
\end{equation}
\end{subequations}
by defining 
\begin{equation}\label{eq:xi}
 \xi=C_d\frac{\mu^2}{B f}   
\end{equation}
Eq. (\ref{eq:rteq_solved}a,b) implies that $r_{t,eq}$ scales with $1/f$ if $r_{RCE,eq}$ scales with $1/f$. Eq. (\ref{eq:rteq_solved}b) also shows that $r_{RCE,eq}>r_{t,eq}. $\footnote{Actually, $r_{RCE,eq}$ would be equivalent to $r_{0,eq}$, the equilibrium radius of vanishing wind, if $\lim\limits_{v_t\to0}\xi v_t^2=0$. A close relation between $r_{RCE,eq}$ and $r_{0,eq}$ is indeed seen in numerical experiments (not shown)}

The time scale for expansion $\tau_{rt}$ in Eq. (\ref{eq:exmodel_general}) is given by 
%
\begin{equation}\label{eq:tau_rt}
\begin{split}
{\tau_{rt}}
&=(-\frac{\partial v}{\partial r})\frac{h_w}{fB}\\
\end{split}
\end{equation}
Here $\tau_{rt}$ is proportional to $1/f$ and $\frac{\partial v}{\partial r}$, meaning that the time scale is larger if $f$ is smaller or the local slope of the wind profile is larger in magnitude. 
Note $\tau_{rt}$ exists independent of the specific parameters for eyewall dynamics, as it depends on $B$ but not $A$.


Conceptually, the model links expansion to the radial velocity $u_t$ induced by the dry entropy imbalance within the TC volume.
A stable equilibrium size ($r_{t,eq}$) independent of time and current size is assumed to exist (Eq. \ref{eq:exmodel_general}).
The following parameters of the model are taken as  constants: 
$\epsilon_{p,ew}$, $\alpha_p$, $h_w$, $f$, $L_v$, $q_{vb}$, $T_{e,lat}$, $T_{e,rad}$, $\Delta s_d$, $\Delta p$, $\rho_i$, and thus $A$, $B$, and $\xi$. 
Doing so simplifies the problem enough to make it analytically tractable.  Simulations also indicate that taking parameters $\epsilon_{p,ew}$, $\alpha_p$, $\Delta s_d$, $\Delta p$ (implicit in Fig. \ref{fig:sdbud_r8_compare_TTPPFCOR}), $\rho_i$ (not shown, also $T_{e,lat}$ and $T_{e,rad}$) as constant is reasonable (see Appendix D)
\footnote{Diagnosed $q_{vb}$ from CTL simulation (Appendix C) increases $\sim 15\%$ during expansion (not shown), but this size-dependence is secondary because the expansion model eventually depends on equilibrium size $r_{t,eq}$. Note also the $q_{vb}$ increase is not explained by a corresponding surface pressure drop, which is only $\sim 2.5\%$.}.
Note a constant $r_{t,eq}$ also implies a constant $M_{ew}/(\rho_wr_t)$ (Eq. \ref{eq:rteq}a).
In this manner, $\mathcal{Q}_{lat}$ is proportional to $r_t$ (Eq. \ref{eq:Qlat}) and $\mathcal{Q}_{rad}$ is proportional to $r_t^2$ (Eq. \ref{eq:Qrad}).
Hence $u_t$ (Eq. \ref{eq:ut_energetics}) monotonically decreases in magnitude with expansion so that TC size will approach an equilibrium. 

If, in addition to $r_{t,eq}$, $\tau_{rt}$ is also time invariant, the solution of Eq. (\ref{eq:exmodel_general}) with initial condition $r_t(t_0)=r_{t0}$ is given by: 
\begin{equation}\label{eq:analy_solution}
    r_t(t)=(r_{t0}-r_{t,eq})e^{-(t-t_0)/\tau_{rt}}+r_{t,eq}
\end{equation}
As $\tau_{rt}$ is positive definite, $r_t$ will exponentially approach the equilibrium size $r_{t,eq}$, where $\tau_{rt}$ is the $e$-folding time scale. 
Moreover, $r_{t,eq}$ is a stable equilibrium, as size approaches $r_{t,eq}$ for $r_{t0}<r_{t,eq}$ (expansion) and $r_{t0}>r_{t,eq}$ (shrinking). Eq. (\ref{eq:analy_solution}) gives an exponential solution, similar to Wang and Toumi (2022) for absolute size, though this solution is exponential in the decay of the distance from equilibrium and hence allows for size to reach an equilibrium value as is known to exist on the $f$-plane.

Up to this point, $\partial v/\partial r$ in $\tau_{rt}$ (Eq. \ref{eq:tau_rt}) is not yet defined analytically, which is needed for a full analytical solution of Eq. (\ref{eq:exmodel_general}). Moreover, $r_{t,eq}$ (Eq. \ref{eq:rteq}) is not yet defined in terms of environmental parameters, which requires an expression for $(M_{ew}/\rho_w)_{eq}$. In the following subsections, we will resolve these issues and obtain a full analytical solution of Eq. (\ref{eq:exmodel_general}).


\subsection{Analytical expression of $\partial v/\partial r$}

Eq. (\ref{eq:rteq_solved}) provides an expression for the equilibrium radii of different wind speeds.
The slope of the equilibrium wind profile $\partial v_t/\partial r_{t,eq}$ can be obtained by taking the derivative of $r_{t,eq}$ with respect to $v_t$ in Eq. (\ref{eq:rteq_solved}) in a fixed environment:
\begin{equation}\label{eq:dvdr_solved}
    (\frac{\partial v_t}{\partial r_{t,eq}})^{-1}=\frac{\partial r_{t,eq}}{\partial v_t}=-\frac{2v_t(\frac{1}{2}v_t\xi'+\xi)r_{t,eq}}{2r_{t,eq}+\xi v_t^2}
\end{equation}
where $\xi'=d\xi/dv_t$. 
Note that $\xi$ is a constant with respect to $r_{t}$ in a given environment when $v_t$ is fixed, but may vary with $v_t$. 
For example, closer to the center (larger $v_t$), absolute vorticity is larger, so $\xi$ should decrease accordingly (though above we have approximated absolute vorticity by $f$).
Here as $\xi'$ should also be a constant with respect to $r_t$ at fixed $v_t$, 
to simplify the math, we take the approximation $\frac{1}{2}v_t\xi'+\xi=\sigma\xi$, where $\sigma$ is set to a constant value of 0.7 (for $v_t=8$ m/s, shown below).
Note $\sigma>0$ is presumed so that the RHS of Eq. (\ref{eq:dvdr_solved}) is negative, corresponding to a TC wind profile in which the azimuthal wind speed decreases with radius.
With this, we rewrite Eq. (\ref{eq:dvdr_solved}) in general by dropping the subscript "eq" and write $\xi$ as $\xi_0$ to mark that it is only associated with $\partial v/\partial r$:
\begin{equation}\label{eq:dvdr_solved2}
\begin{split}
    (\frac{\partial v}{\partial r})^{-1}\bigg|_{v=v_t}=\frac{\partial r}{\partial v}\bigg|_{v=v_t}
    &=-\frac{2r_tv_t\sigma\xi_0}{2r_t+\xi_0 v_t^2},
\end{split}
\end{equation}
The assumption implicitly made to move from Eq. (\ref{eq:dvdr_solved}) to Eq. (\ref{eq:dvdr_solved2}) is that $(\frac{\partial v}{\partial r})|_{v=v_t}$ at given $r_t$ and $v_t$ for slowly-evolving wind fields can be approximated by equilibrium values. This assumption follows the fairly nice performance of E04 model, which is derived for steady-state, of matching observed TC outer wind profiles for storms that are not necessarily in steady-state (Chavas et al. 2015). This assumption will be shown to work nicely in Sec. 3. For our second assumption that $\sigma$ is set to be a fixed constant, this choice assumes that $\xi_0$ is proportional to $C_d/f$ in the same manner as $\xi$ (Eq. \ref{eq:xi}), but it does not depend on $\mu^2/B$ (otherwise we would have to find the useful $\sigma$ every time when $\mu^2/B$ changes). 
In practice, we set 
\begin{equation}
    \xi_0=b\frac{C_d}{f}
\end{equation}
where $b$ is a constant, equal to the $\mu^2/B$ in the baseline environment, to be defined in the next paragraph and Sec. 3.
Eq. (\ref{eq:dvdr_solved2}) provides an analytical expression of $\partial v/\partial r$.  
Compared to E04 model, Eq. (\ref{eq:dvdr_solved2}) does not need numerical integration. Additional discussion of the properties of Eq. (\ref{eq:dvdr_solved2}) is provided in Appendix A. In the next subsection, we will substitute Eq. (\ref{eq:dvdr_solved2}) into Eq. (\ref{eq:tau_rt}) to yield an analytical solution of Eq. (\ref{eq:exmodel_general}).

\begin{figure}[t]
 \noindent\includegraphics[width=0.5\textwidth,angle=0]{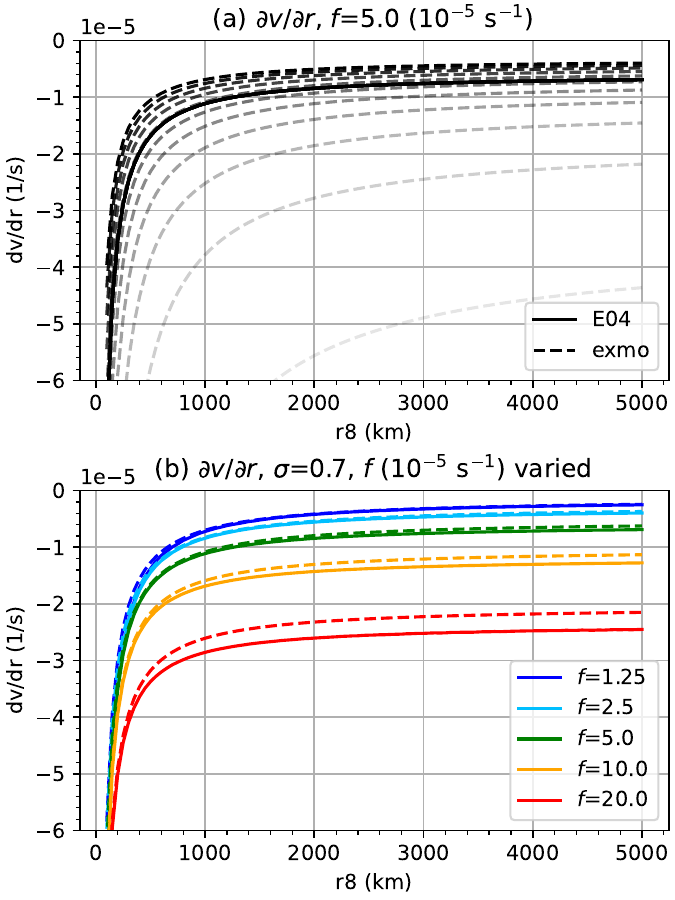}\\
 \caption{(a) $\partial v/\partial r$ at $r_8$ from E04 model (solid) and expansion model Eq. (\ref{eq:dvdr_solved2}) (dashed) with $\sigma$ varied from 0.1 to 1.1 (light to dark, with an interval of 0.1) at $f=5\times10^{-5}$ s$^{-1}$. (b) as in (a), but with $\sigma=0.7$  and with different $f$ ($10^{-5}$ s$^{-1}$, see legend). See text for parameter settings. 
 }\label{fig:exmo_wind}
\end{figure}

Now we demonstrate that $\sigma=0.7$ is useful for $\partial v/\partial r$ (Eq. \ref{eq:dvdr_solved2}) at $v_t=8$ m/s (this specific $v_t$ will be used for evaluation of the model in Sec. 3-4). 
We define a baseline environment of $\xi_0=35105$ s$^2$/m  with $f=5\times10^{-5}$ s$^{-1}$ and $C_d=0.0015$ for demonstration (note a complete parameter setting in this baseline environment is given in Sec. 3 below).  
Thus, we can write $\xi_0=35105\times(5\times10^{-5}/0.0015)\times C_d/f=1170.17\times C_d/f$ in case $C_d/f$ varies (so $b=1170.17$).
The quantity $\partial v/\partial r$ at $r_8$ from E04 model (solid) and in Eq. (\ref{eq:dvdr_solved2}) (dashed) in the baseline environment are shown in Fig. \ref{fig:exmo_wind}a. 
Parameter $\sigma$ is varied from 0.1 to 1.1 to show the sensitivity of Eq. (\ref{eq:dvdr_solved2}) to this quantity. Indeed, Eq. (\ref{eq:dvdr_solved2}) with $\sigma = 0.7$ does very well in reproducing $\partial v/\partial r$ for any value of $r_8$ and over a wide range of values of $f$, compared to E04 model. 
Thus, we take $\sigma=0.7$ below.

Combining Eqs. (\ref{eq:tau_rt}) and (\ref{eq:dvdr_solved2}) gives a final expression for $\tau_{rt}$:
\begin{equation}\label{eq:tau_rt_2}
    \tau_{rt}=\frac{2r_t+\xi_0v_t^2}{2r_tv_t\sigma\xi_0}\cdot\frac{h_w}{fB}
\end{equation}

\subsection{Analytical solution for size evolution}

Substituting Eq. (\ref{eq:tau_rt_2}) into Eq. (\ref{eq:exmodel_general}) yields:
\begin{equation}\label{eq:drtdt}
    \frac{dr_t}{dt}=2\frac{fB}{h_w}\sigma\xi_0 v_t\frac{r_t(r_{t,eq}-r_t)}{2r_t+\xi_0 v_t^2},
\end{equation}
which gives an explicit form of the expansion rate ($dr_t/dt>0$ when $0<r_t<r_{t,eq}$ and $dr_t/dt<0$ when $r_t>r_{t,eq}$). Eq. (\ref{eq:drtdt}) again indicates that $r_{t,eq}$ is a stable equilibrium,
and $\frac{dr_t}{dt}=0$ when $r_t=r_{t,eq}$ (and for $r_t=0$). 
The maximum expansion rate occurs at a size given by:
\begin{equation}\label{eq:rtexp}
    r_{t,expmax}=\frac{-\xi_0 v_t^2+\sqrt{\xi_0^2v_t^4+2r_{t,eq}\xi_0 v_t^2}}{2},
\end{equation}
, where it is seen $0<r_{t,expmax}<r_{t,eq}/2$. 
When $r_{t}>r_{t,eq}$, $\frac{\partial}{\partial r_t}(dr_t/dt)<0$, meaning that size shrinks faster towards $r_{t,eq}$ when $r_t$ is farther from $r_{t,eq}$.

Solving Eq. (\ref{eq:drtdt}) with initial condition $r_t=r_{t0}$ at $t=t_0$ gives:
\begin{equation}\label{eq:exmo_analy}
\begin{split}
    &\quad t-t_0\\
    &=\frac{1}{2\xi_0 v_t\frac{fB\sigma}{h_w}}[-(2+\frac{\xi_0 v_t^2}{r_{t,eq}})\ln(\frac{r_{t,eq}-r_t}{r_{t,eq}-r_{t0}})+\frac{\xi_0 v_t^2}{r_{t,eq}}\ln(\frac{r_t}{r_{t0}})], \\
    &\quad\quad\quad\quad\quad\quad\quad\quad \text{$r_t>0$ and  $r_t\neq r_{t,eq}$}
\end{split}
\end{equation}
Eq. (\ref{eq:exmo_analy}) is the analytical solution of the full size-expansion model in Sec. 2a. The solution is expressed by time $t$ as a function of $r_t$, which is an implicit function of $t$; an analytic solution for $r_t(t)$ is not tractable. The input parameters are all external or environmentally defined (presently $r_{t,eq}$ can be either external or environmentally defined by TC PS). A method for determining $r_{t,eq}$ (Eq. \ref{eq:rteq_solved}b) from environmental parameters is provided next.

\subsection{Formulation for updraft mass flux}

An environmentally defined $r_{t,eq}$ will be obtained through Eqs. (\ref{eq:r0_eq}) and (\ref{eq:rteq_solved}b) if $(M_{ew}/\rho_w)_{eq}$ is environmentally defined. In this subsection, we parameterize $(M_{ew}/\rho_w)_{eq}$ by using a combination of theory and empirical estimation based on numerical simulation results.

%

The parameterization may be derived directly from mass continuity: eyewall updraft mass flux is balanced by a constant subsidence velocity, which is usually assumed to be driven by radiative cooling (e.g. E04). The streamfunction is given by
$\frac{\partial \psi}{\partial r}=2\pi\rho_drw$, with $\rho_d$ the dry air density and $w$ the vertical velocity.
Integrating radially over the subsidence region at the altitude of $h_w$ yields
\begin{equation}
r_{\psi_0}^2\approx\frac{\psi_{max}}{\pi\rho_w w_{cool}}+r_{\psi_{max}}^2 ,
\end{equation}
where $r_{\psi_0}$ and $r_{\psi_{max}}$ are the radii of $\psi=0$ and maximum $\psi$ (or $\psi_{max}$) at $h_w$, respectively, and $w_{cool}$ is the environmental clear-air subsidence velocity (positive downward). 
The inner radius term $r_{\psi_{max}}^2$ may be neglected as it is more than an order of magnitude smaller than $r_{\psi_0}^2$. 
Hence
\begin{equation}
    \psi_{max}\approx\pi\rho_w w_{cool}r_0^2
\end{equation}
where $r_0$, the radius of vanishing wind, should be equivalent to $r_{\psi_{0}}$ in E04.
%
TCPS shows that equilibrium $r_{0}$ (or $r_{0,eq}$) scales with $V_{Carnot}/f$, which does not depend on $C_d$. This is consistent with
\begin{equation}
    r_{0,eq}\propto C_d^{0.5}V_p/f
\end{equation}
Following our assumption that most of the upward mass flux occurs within the eyewall so that $M_{ew}\approx\psi_{max}$, thus, we propose that
\begin{equation}\label{eq:mew_prop_CdVp_f}
\sqrt{(M_{ew}/\rho_w)_{eq}} \propto \sqrt{\pi w_{cool}}C_d^\nu V_p/f   
\end{equation}
, a relationship we test with numerical simulations, with $\nu$ a constant coefficient. We use a generalized $V_p$, which is defined following Chavas and Emanuel (2014), based on Bister and Emanuel (1998\nocite{Bister_Emanuel_1998}) and Rousseau-Rizzi and Emanuel (2019\nocite{Rousseau-Rizzi_Emanuel_2019}):
\begin{equation}\label{eq:vp}
V_p^2=\frac{C_k}{C_d}\frac{T_{SST}-T_{tpp}}{T_{tpp}}(k_0^*-k),
\end{equation}
where $C_k$ and $C_d$ are surface exchange coefficients for enthalpy and momentum, $T_{SST}$ the sea surface temperature, $T_{tpp}$ the tropopause temperature, $k_0^*$ the saturation enthalpy at $T_{SST}$ and $k$ the actual enthalpy of near surface air.

\begin{figure}[t]
 \noindent\includegraphics[width=20pc,angle=0]{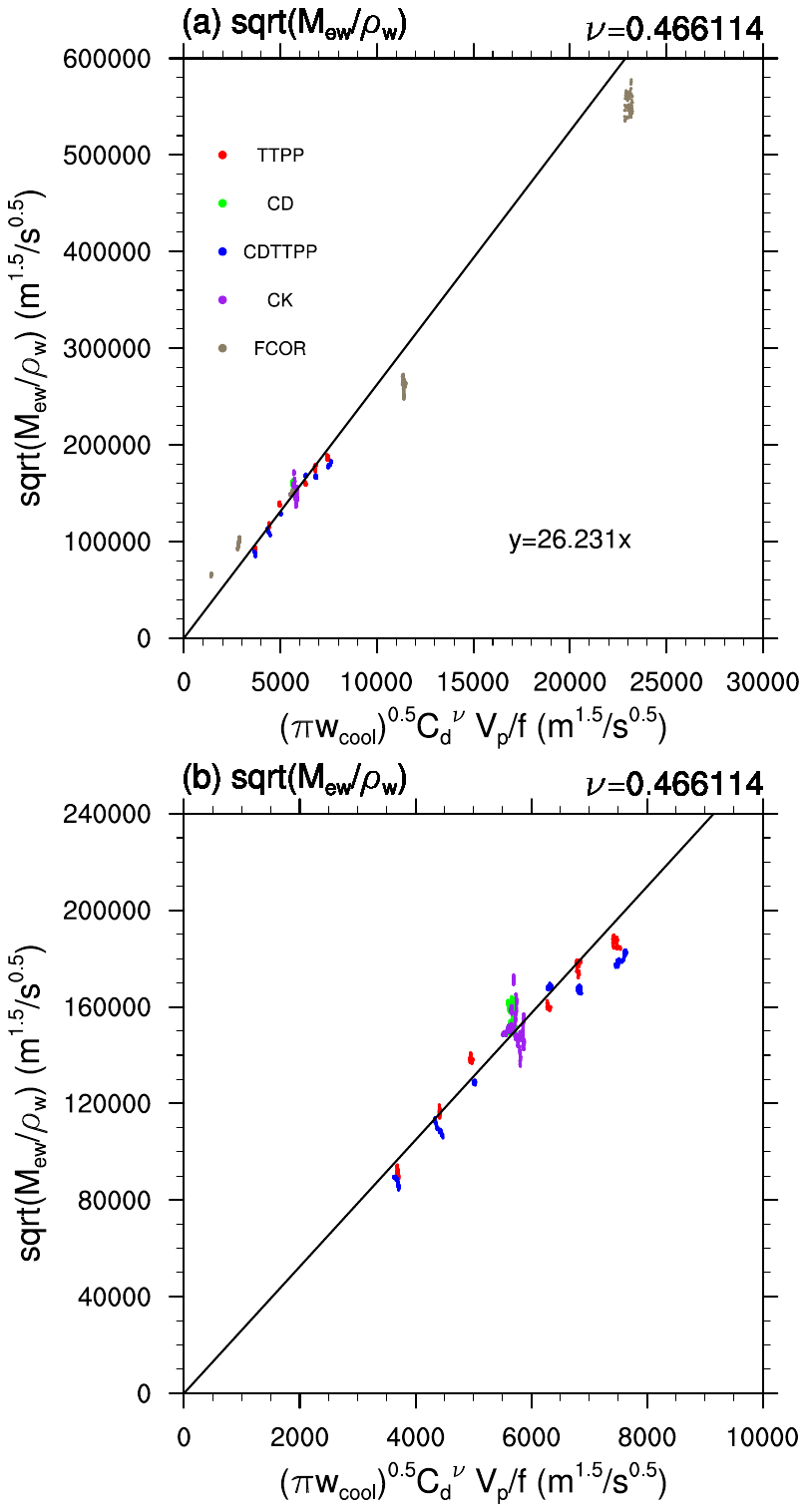}\\
 \caption{(a) $\sqrt{(M_{ew}/\rho_w)_{eq}}$ (m$^{1.5}$/s$^{0.5}$) as a function of $\sqrt{\pi w_{cool}}C_d^{\nu}V_{p}/f$ (m$^{1.5}$/s$^{0.5}$) in TTPP (red), CD (green), CDTTPP (blue), CK (purple) and FCOR (grey) during equilibrium periods. Data of TTPP, CD, CDTTPP and CK are used to determine $\nu$ by linear regression. Fitted $\nu$ is shown on the upper right of the plot. Data are first processed by a 120-h running average. The black line visualizes the equation shown in the figure. In the equation, $y=\sqrt{(M_{ew}/\rho_w)_{eq}}$ and $x=\sqrt{\pi w_{cool}}C_d^{\nu}V_{p}/f$. See Appendix C for experimental design. (b) as in (a) but zoomed in without FCOR. Equilibrium periods for TTPP and FCOR are defined in Sec. 3b below, and those for CD, CDTTPP and CK are same as TTPP.
 }\label{fig:fit_M_ew_ov_rho}
\end{figure}

The exact relationship is not known, and thus we seek the relation in Eq. (\ref{eq:mew_prop_CdVp_f}) via linear regression from equilibrium states of simulated TCs (see Appendix C). There is a tight linear relationship between the two quantities in Eq. (\ref{eq:mew_prop_CdVp_f}). We estimate the coefficient based on the linear fit to the experiment sets varying $T_{tpp}$, $C_d$, and $C_k$ to avoid overfitting to the  experiments varying $f$, whose slope deviates slightly, but the result holds reasonably well for those experiments too.
The result is shown in Fig. \ref{fig:fit_M_ew_ov_rho}. A best-fit estimate of $\nu=0.47$ is obtained. As $C_d^{-0.5}$ exists in $V_p$, then the fitting suggests $C_d$ has nearly zero effect on $(M_{ew}/\rho_w)_{eq}$, consistent with the finding in Wang et al. (2022). As a final result of the fitting, we have
\begin{equation}\label{eq:fitted_Mewovrho}
    \sqrt{(\frac{M_{ew}}{\rho_w})_{eq}}=26.2\sqrt{\pi w_{cool}}C_d^{0.47}V_p/f
\end{equation}

The environmentally defined $r_{t,eq}$ is obtained by substituting Eq. (\ref{eq:fitted_Mewovrho}) into Eq. (\ref{eq:r0_eq}) and substituting the resulting $r_{RCE,eq}$ into Eq. (\ref{eq:rteq_solved}b). As discussed in Sec. 2b, $r_{t,eq}$ will scale with $1/f$ and additionally increase with $C_d^{0.47}V_p$. The expansion model is now capable of being fully predictive and analytic.

\subsection{Model summary and implementation}

To summarize, this section has proposed a model for the expansion of TC size in which expansion is driven by latent heating, which is dominated by heating in the eyewall, and suppressed by radiative cooling. The model can be fully predictive and analytic if there is an environmentally defined size-independent $r_{t,eq}$. 
The steps to put the model into practice are as follows:
\begin{itemize}
    \item Assuming known $r_{t,eq}$, the evolution of outer radius $r_t$ is given by Eq. (\ref{eq:exmo_analy}). Parameter $B$ given in Eq. (\ref{eq:AB}b). For $v_t=8$ m/s, $\xi_0=35105$ s$^2$/m for $C_d=0.0015$ and $f=5\times10^{-5}$ s$^{-1}$ (or $\xi_0=1170.17\times C_d/f$ otherwise) and $\sigma=0.7$.
    \item Equilibrium size $r_{t,eq}$ is predicted from environmental parameters by using Eq. (\ref{eq:fitted_Mewovrho}) for $(\frac{M_{ew}}{\rho_w})_{eq}$, plugging the result into Eq. (\ref{eq:r0_eq}) to calculate $r_{RCE,eq}$ and plugging the result into Eq. (\ref{eq:rteq_solved}) for $r_{t,eq}$. 
\end{itemize}
Once given $r_{t,eq}$, $\mathcal{Q}_{lat}$ as a proportional function of $r_t$ is calculated through Eqs. (\ref{eq:rteq_solved},\ref{eq:r0_eq},\ref{eq:Qlat}), or directly by Eq. (\ref{eq:rteq}b). Radial velocity $u_t$ is given by Eq. (\ref{eq:ut_energetics}); local tangential wind acceleration $\partial v/\partial t$ (at $r=r_t$) is given by Eq. (\ref{eq:vten2}).

In Sec. 3a, we examine the basic properties of the analytical solution for the evolution of storm size as well as the underlying physical processes of the model. In Sec. 3b-d, we use numerical simulations to test the model predictions of both expansion rate (Sec. 3b) and $r_{t,eq}$ (Sec. 3c-d).
Sec. 4a explores the physical meaning of $r_{t,eq}$. Sec. 4b examines the model's representation of the dependence of inflow velocity, local tangential wind spinup and size expansion rate on latent heating rate.  Sec. 4c examines the sensitivity of the model to $\Delta s_d$.

\section{Evaluation of the model}

We next discuss the basic behavior of the expansion model solution and compare it with numerical simulation experiments. 
We take $v_t=8$ m/s as our outer size wind speed, and hence we use $r_8$ in lieu of $r_t$ at times.
We define an idealized baseline environment for analysis with parameter values representative of tropical cyclones in the present day tropical atmosphere. We set as constants: $f=5\times10^{-5}$ s$^{-1}$; $Q_{cool}=1$ K/day; $h_w=2.5$ km (depth that captures the majority of the lateral inflow mass flux); $\rho_i=1.1$ kg/m$^3$; $C_d=C_k=0.0015$; $\mu=0.92$ (to match CM1 simulations, where a surface gustiness has been added, described in Appendix C); $L_v=2.501\times10^6$ J/kg; $\epsilon_{p,ew}=1$, $\alpha_p=0.8$ (indicating that the eyewall dominates net latent heating in a TC, estimated from simulations in Appendix C);
$\sigma=0.7$.
$V_p$ (Eq. \ref{eq:vp}) is defined with $T_{SST}=300$ K, $T_{tpp}=200$ K, air-sea temperature difference 0 K so that surface air temperature $T_s=300$ K, environmental near-surface relative humidity $\mathcal{H}_e=0.87$, environmental surface pressure $p_s=1015$ hPa,  which collectively yields $V_p=60.4$ m/s. These values, including the final value of $V_p$, are close to that in CTL simulation (Fig.\ref{fig:rteq_Mewovrho_eq_TTPP_FCOR} in Sec. 3b).
We further set $q_{vb} =q_{vs}^*= 0.022$ kg/kg, with $q_{vs}^*$ the environmental saturation mixing ratio of at the surface air temperature\footnote{This can be a $\sim 20\%$ overestimate of $q_{vb}$ as it does not account for the vertical profile of boundary-layer $q_v$. This would lead to a $\sim 200$ km overestimate of $r_{t,eq}$ in Sec. 4c. Taking a vertical average from surface to 2 km of altitude (approximately the inflow depth associated with $M_{ew}$) of the analytical saturation mixing ratio $q_{v}^*$ profile of Romps (2016\nocite{Romps2016}, his Eqs. 8 and 11) appears to resolve this issue, though here we keep it simple and not adopting the Romps (2016) model. }. 
Parameter $\Delta s_d$ is set to $L_vq_{vs}^*/T_{s}=187.1$ J/K/kg, and we set 
$T_{e,rad}=T_{e,lat}=T_s$ without loss of generality \footnote{In simulations,  $T_{e,lat}$ and $T_{e,rad}$ are about $275$ K, though here we avoid specifying these values based on simulations for simplicity. Our approach yields $T_{e,lat}\Delta s_d=T_{e,rad}\Delta s_d=L_vq_{vb}$, i.e. a characteristic difference of dry static energy between tropopause and surface. This $\sim10\%$ overestimation $T_{e,lat}$ and $T_{e,rad}$ would induce a $\sim10\%$ overestimation of $\xi$, and a $\sim 55$ km underestimate of $r_{t,eq}$ in Sec. 4c. }.
Tropopause pressure $p_t$ is set to 100 hPa (for $\mathcal{Q}_{rad}$, Eq. \ref{eq:Qrad}, which gives 89.3 W/m$^2$). 
These settings yield $A=0.16$, $B=0.0007$ m/s, $\xi=35105$ s$^2$/m and we set $\xi_0\equiv\xi$ through out, except in Sec. 4c. Finally, we set $w_{cool}=0.003$ m/s, which will only be used to calculate $r_{t,eq}$ from $(M_{ew}/\rho_w)_{ew}$ (Eq. \ref{eq:fitted_Mewovrho}) in Sec. 4c below. Unless otherwise noted, these values are held constant throughout so that the use of the analytic model in as simple a setup as possible.

These constants are complete for the fully predictive size expansion model and will be used for tests below unless otherwise noted. In particular, we will in times arbitrarily set $r_{t,eq}$ for certain examinations in Sec. 3-4, which will also be noted.

\subsection{Basic behavior of TC size evolution model}

In this subsection, we examine the basic behavior of our TC size evolution model and the underlying physical processes.

\begin{figure}[t]
 \noindent\includegraphics[width=20pc,angle=0]{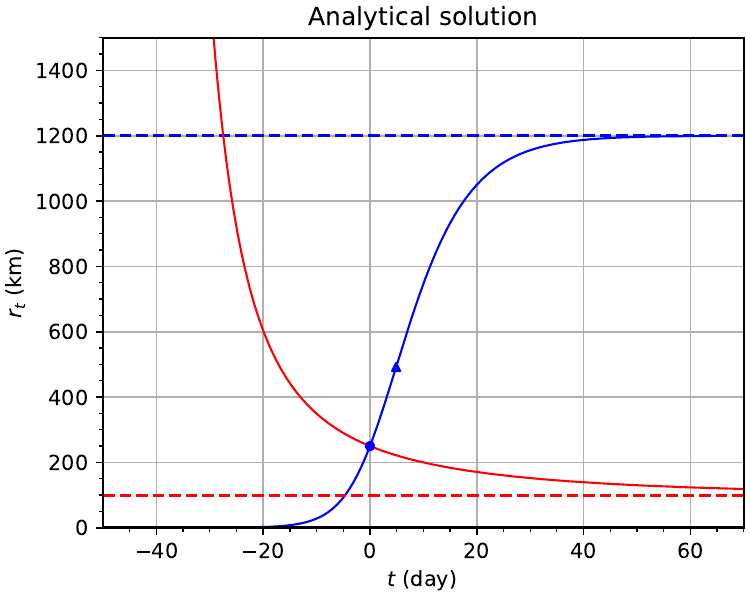}\\
 \caption{Analytical solution of size evolution ($r_t$, km vs. $t$, day, solid lines) in Eq. (\ref{eq:exmo_analy}) for two cases: $r_t$ expanding towards a larger $r_{t,eq}$ (blue) and $r_t$ shrinking towards a smaller $r_{t,eq}$ (red). Horizontal dashed lines mark $r_{t,eq}$. Dots mark initial condition $t_0$ and $r_{t0}$. Triangle mark the location ($r_{t,expmax}$) of maximum expansion rate. 
 }\label{fig:exmo_analy}
\end{figure}

\begin{figure}[t]
 \noindent\includegraphics[width=0.4\textwidth,angle=0]{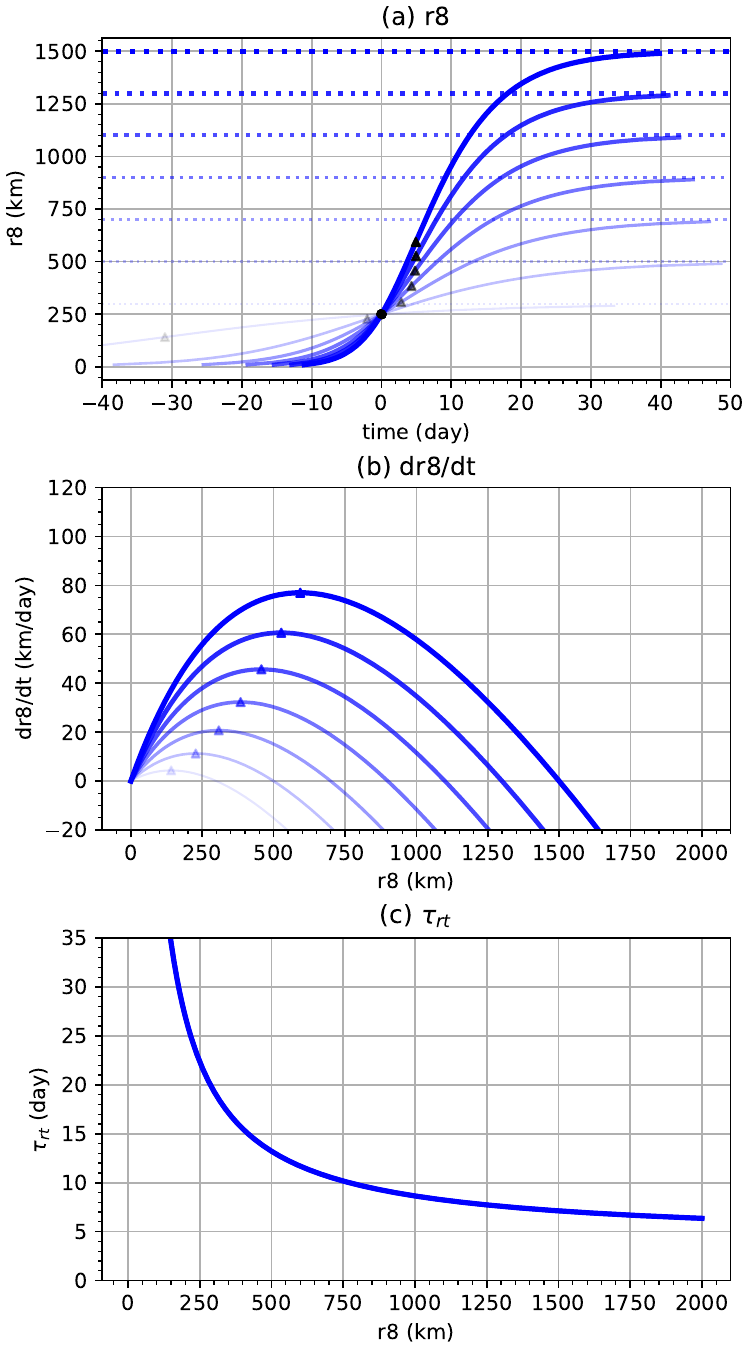}\\
 \caption{Idealized expansion model prediction. (a) Time evolution of $r_8$ (solid and dashed) with $r_{t,eq}$ (dotted); (b) $dr_8/dt$ (km/day) as a function of $r_8$ (km); (c) $\tau_{rt}$ (day) as a function of $r_8$ (km); In (a)-(c), thicker and more opaque lines mark higher values of $r_{t,eq}$. Dot marks initial condition and triangles mark the location of maximum expansion rate.
 }\label{fig:model_ideal_1}
\end{figure}

First, the basic evolution of size predicted from our model (Eq. \ref{eq:exmo_analy}) is shown in Fig. \ref{fig:exmo_analy}. 
Two representative cases are shown from an initial size of $r_{t0} = 250$ km at initial time $t_0 = 0$ day: expansion towards a larger equilibrium size of $r_{t,eq}=1200$ km (blue curve), and shrinking towards a smaller equilibrium size of $r_{t,eq}=100$ km (red curve). 
For both cases, the model predicts a reasonable time scale of 10-20 days (note $\tau_{rt}=22.4, 13.2, 10.2, 7.9$ days at $r_8=250, 500, 750, 1200$ km, respectively, Fig. \ref{fig:model_ideal_1}c). 
The rate of expansion/shrinking vanishes as size approaches its equilibrium ($r_{t,eq}$ and 0).  The maximum expansion rate  occurs during the first half of expansion process at a radius $r_{t,expmax}$ of approximately 500 km (Eq. \ref{eq:rtexp}). 

Next, we show the solution for expansion towards a range of equilibrium sizes ($r_{t,eq}$). 
The size evolution $r_t(t)$, expansion rate $dr_t/dt$, time scale $\tau_{rt}$ are shown in Fig. \ref{fig:model_ideal_1}.
Radius $r_8$ increases with time and approaches $r_{t,eq}$ after day 20. 
Larger expansion rate corresponds to larger $r_{t,eq}$ (Fig. \ref{fig:model_ideal_1}b). This is because the time scale $\tau_{rt}$ is the same across all experiments (Eqs. \ref{eq:exmodel_general}, \ref{eq:tau_rt} and \ref{eq:dvdr_solved2}, Fig. \ref{fig:model_ideal_1}c), as this quantity is a function of size alone in this example. Time scale $\tau_{rt}$ monotonically decreases with size, with a first rapid decrease when $r_8<500$ km and slowly decrease afterwards. Physically this is because E04 model is flatter (smaller slope) for larger storm (longer tail) so from Eq. (\ref{eq:tau_rt}) $\tau_{rt}$ becomes smaller too. 
Note $\tau_{rt}$ is greater than 15 days when the TC is small ($r_8\approx 400$ km) and decreases below 10 days as size approaches $r_{t,eq}$. 
The variation of $\tau_{rt}$ is determined by the variation of $\partial v/\partial r$ at $r_8$ (Eq. \ref{eq:tau_rt}, Fig. \ref{fig:exmo_wind}).
Correspondingly, the expansion rate peaks at tens of kilometers per day (Fig. \ref{fig:model_ideal_1}b), a similar order of magnitude to that seen in observations (Schenkel et al. 2023). Finally, the radius of maximum expansion rate increases with $r_{t,eq}$, following Eq. (\ref{eq:rtexp}).   


\begin{figure}[t]
 \noindent\includegraphics[width=0.4\textwidth,angle=0]{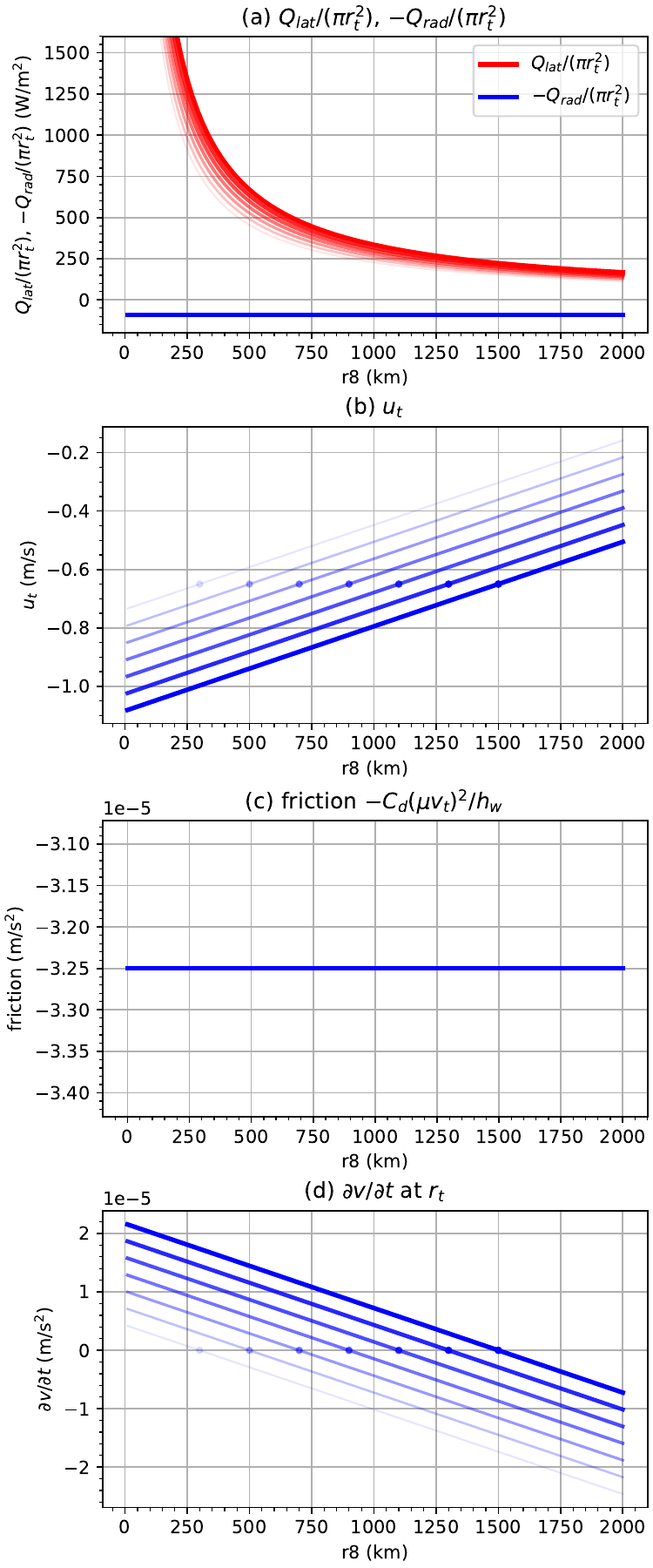}\\
 \caption{Same as Fig. \ref{fig:model_ideal_1}, but for (a) latent heating per unit area $\mathcal{Q}_{lat}/(\pi r_t^2)$ (W/m$^2$, red) and radiative cooling per unit area $-\mathcal{Q}_{rad}/(\pi r_t^2)$ (W/m$^2$, blue), (b) radial velocity $u_t$ (m/s),  (c) $-C_d(\mu v_t)^2/h_w$ (m/s$^2$), (d) local spinup rate $\partial v/\partial t$ (m/s$^2$) at $r_t$,  The dots in (b) and (d) mark equilibrium.
 }\label{fig:model_ideal_2}
\end{figure}

We next explore physically why the TC expands in the first place and why it eventually reaches equilibrium, following the conceptual diagram in Fig. \ref{fig:schematic}. We examine the budget terms in the expansion equation (Eq. \ref{eq:drtdt_0}) and the equation for the dependence of the inflow velocity on latent heating and radiative cooling (Eq. \ref{eq:ut_energetics}a).
The five underlying processes/terms in the schematic (Fig. \ref{fig:schematic}) are: the latent heating per unit area $\mathcal{Q}_{lat}/(\pi r_t^2)$, radiative cooling per unit area $\mathcal{Q}_{rad}/(\pi r_t^2)$, $u_t$, turbulent friction $-C_d(\mu v_t)^2/h_w$, $\partial v/\partial t$ at $r_t$. Our model prediction of each of these terms is shown in Fig. \ref{fig:model_ideal_2}.

During the expansion stage latent heating is significantly larger than radiative cooling. The latent heating rate can exceed 900 W/m$^2$ when the TC is small ($r_8\approx250$ km), which is also supported by the simulations in Appendix C with $f=5\times10^{-5}$ s$^{-1}$ (not shown). 
This amount of heating would induce $\sim8$ K/day temperature increase of the atmosphere (assuming constant pressure) without lateral energy/entropy exchange, whereas the actual average temperature change rate is of the order $10^{-1}$ K/day in a TC (estimated from the simulation with $T_{tpp}=200$ K and $f=5\times10^{-5}$ s$^{-1}$ in Appendix C).
Thus the overturning circulation is needed to export excess latent heating in the TC by exporting higher-entropy air aloft and importing lower-entropy air at low levels. The induced low-level inflow by this overturning circulation may be strong enough so that local spinup at $r_t$ is achieved and TC starts to expand. 
Quantity $u_t$ linearly increases with $r_t$ (Eq. \ref{eq:ut_energetics}) with an equilibrium value about -0.65 m/s (Fig. \ref{fig:model_ideal_2}b), which corresponds to zero local spinup.
Friction at $r_t$ is a constant by design at a value of  $\sim -3.25\times10^{-5}$ m/s$^2$ (Fig. \ref{fig:model_ideal_2}c).  As a result, $\partial v/\partial t$ at $r_t$ linearly decreases with size (Eq. \ref{eq:vten2}), such that an equilibrium is guaranteed. 

It follows from Eq. (\ref{eq:rteq}) that larger $r_{t,eq}$ corresponds to larger $\mathcal{Q}_{lat}$ (Fig. \ref{fig:model_ideal_2}a), and thus the quantity $\mathcal{Q}_{lat}/(\pi r_t^2) \sim 1/r_t$. Meanwhile, the quantity $\mathcal{Q}_{rad}/(\pi r_t^2)$ is a constant $\sim 89$ W/m$^2$. Thus, there exists a TC size at which there is zero net heating in the TC, and thus the expansion rate must vanish before this radius in reached. 
At equilibrium itself, the area-integrated latent heating inside of $r_{t,eq}$ still slightly exceeds radiative cooling because nonzero surface friction also exists (Eqs. \ref{eq:drtdt_0}, \ref{eq:ut_energetics}).

Finally, taking $\partial v/\partial t$ (Fig. \ref{fig:model_ideal_2}d) and $\partial v/\partial r$ (Fig. \ref{fig:exmo_wind}) together, the expansion rate (Fig. \ref{fig:model_ideal_1}c) peaks in the middle of expansion rather than the beginning because of the larger slope of the wind profile when the TC is small.

\subsection{
Evaluating modeled expansion assuming known equilibrium size
}

In this subsection, we test our model's prediction for the time-dependent evolution of size against numerical simulations for the case where $r_{t,eq}$ is known. 
To do this,
we set $r_{t,eq}$ constant and equal to the equilibrium size of the simulated TC in each experiment to evaluate how well the analytical expansion model solution (Eq. \ref{eq:exmo_analy}) can capture the first-order structure of expansion and its variations across experiments. We define $\tau_{rt}$ using the parameters of the base environment in Sec. 3a. 
Two sets of numerical simulations (see Appendix C) are taken for comparison: one set varying the tropopause tempearture $T_{tpp}$ (TTPP), which modulates the potential intensity, the other varying $f$ (FCOR). 
CTL experiment is defined at $T_{tpp}=200$ K and $f=5\times 10^{-5}$ s$^{-1}$, corresponding to TCs on real earth.

\subsubsection{Comparison with TTPP}
\begin{figure}[t]
 \noindent\includegraphics[width=17pc,angle=0]{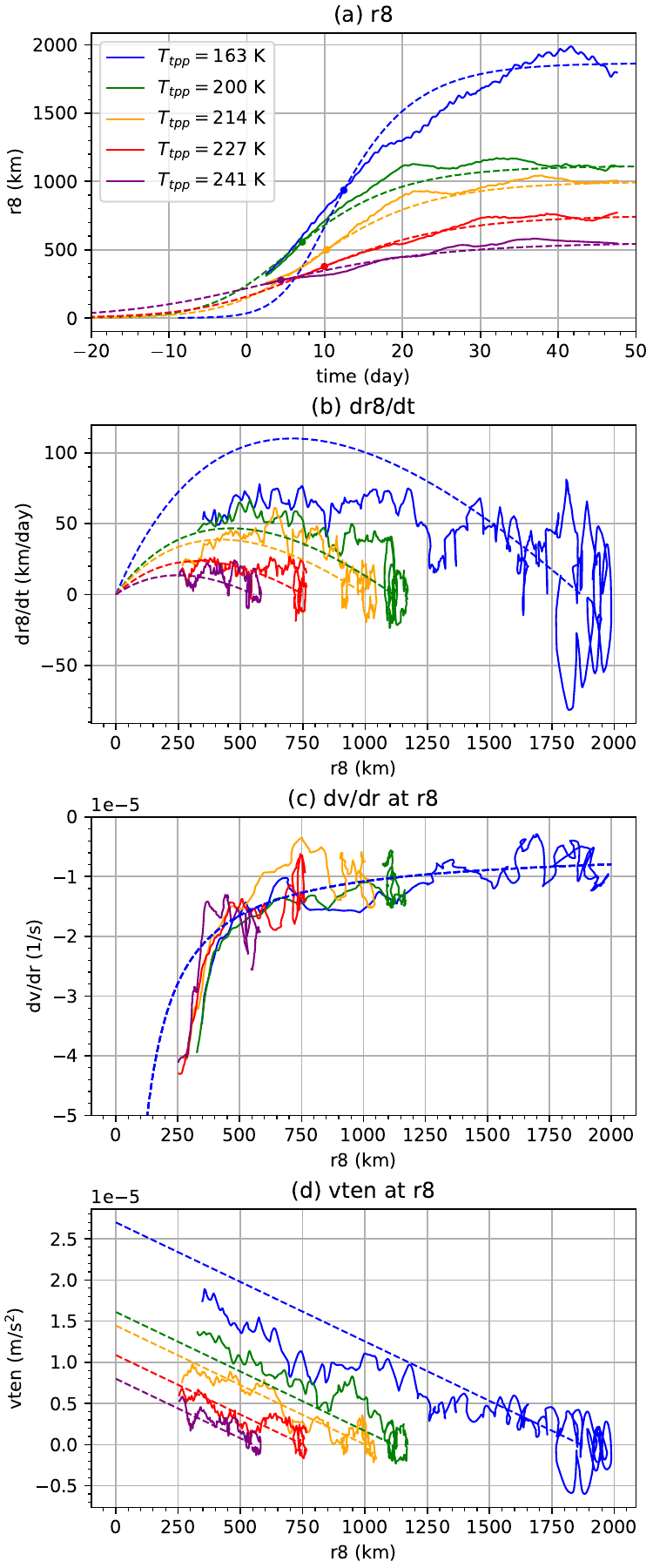}\\
 \caption{(a) Temporal evolution of 120-h running averaged $r_8$ (km) in TTPP (solid, cases with $T_{tpp}=241, 227, 214,200,163$ K are shown) and analytical prediction of the expansion model (dashed) taking $r_{t,eq}$ equal to equilibrium sizes of TTPP (see text). Dot mark initial condition for expansion model. (b) The corresponding expansion rate $dr_8/dt$ (km/day) in TTPP (solid, shown as the 24-h expansion rate of the 120-h running averaged $r_8$) and in expansion model (dashed). (c) $\partial v/\partial r$ (1/s) at $r_8$ in TTPP (solid) and in expansion model (dashed, Eq. \ref{eq:dvdr_solved2}). (d) $\partial v/\partial t$ (m/s$^2$) at $r_8$ in TTPP (solid) and in expansion model (dashed, Eq. \ref{eq:vten2}). Warmer color means higher $T_{tpp}$.
 }\label{fig:TTPP_size_comp}
\end{figure}

The size evolutions across experiments for TTPP are shown in Fig. \ref{fig:TTPP_size_comp}a (solid line). 
Cases with $T_{tpp}=174, 187$ K are neglected to show (here and throughout, except in Sec. 4b) because their size evolution is very similar to $T_{tpp}=163$ K. It is seen that TC size increases with time for about 20 days and approaches an equilibrium, similar to the qualitative behavior of ideal expansion model prediction in Sec. 3a. 
We take the average of $r_8$ during day 40-50 in TTPP as $r_{t,eq}$ for our expansion model; 
in this manner, the expansion model only needs $\tau_{rt}$, which is the same as in Sec. 3a (Fig. \ref{fig:model_ideal_1}b) as determined by the same parameters as in Sec. 3a.

To compare theory and simulation, we set $r_{t0}$ and $t_0$ in our expansion model (Eq. \ref{eq:exmo_analy}) to be the first output $r_8$ above $r_{t,eq}/2$ and the corresponding time in simulations, respectively, for all cases. 
This approach was used to compare intensification theory against simulations in Ramsay et al. (2020\nocite{Ramsay_Singh_Chavas_2020}).
The analytical model predictions of size evolution (Eq. \ref{eq:exmo_analy}) are shown in dashed lines (Fig. \ref{fig:TTPP_size_comp}a). 
Overall, there is a  very good match between TTPP and expansion model prediction.
The lone case that matches a bit less well is the 163 K case, which expands more slowly than the theory predicts throughout expansion. 
The comparison of the corresponding expansion rate $dr_8/dt$ is shown in Fig. \ref{fig:TTPP_size_comp}b. 
The expansion model predicts higher expansion rate (Eq. \ref{eq:drtdt}) with higher $r_{t,eq}$ (thus lower $T_{tpp}$), consistent with the results from TTPP. Moreover, the expansion rate peaks in the early half of expansion in both the expansion and in the TTPP experiments.

The model prediction for $\partial v/\partial r$ at $r_8$ (Eq. \ref{eq:dvdr_solved2}) compares well with those in TTPP (Fig. \ref{fig:TTPP_size_comp}c). The model underestimates the magnitude when $r_8$ is less than 400 km. The model prediction for local spinup rate $\partial v/\partial t$ at $r_8$ also compares well with experiments TTPP (Fig. \ref{fig:TTPP_size_comp}d). The expansion model correctly predicts the qualitative behaviour of $\partial v/\partial t$, with larger values for larger $r_{t,eq}$ and a quasi-linear decrease with expansion. Finally, the dependences of $\partial v/\partial t$ and $\partial v/\partial r$ on $r_8$ in TTPP further confirms that expansion rate peaks in the middle of expansion because the magnitude of $\partial v/\partial r$ is larger when the TC is small.

\subsubsection{Comaprison with FCOR}

\begin{figure}[t]
 \noindent\includegraphics[width=17pc,angle=0]{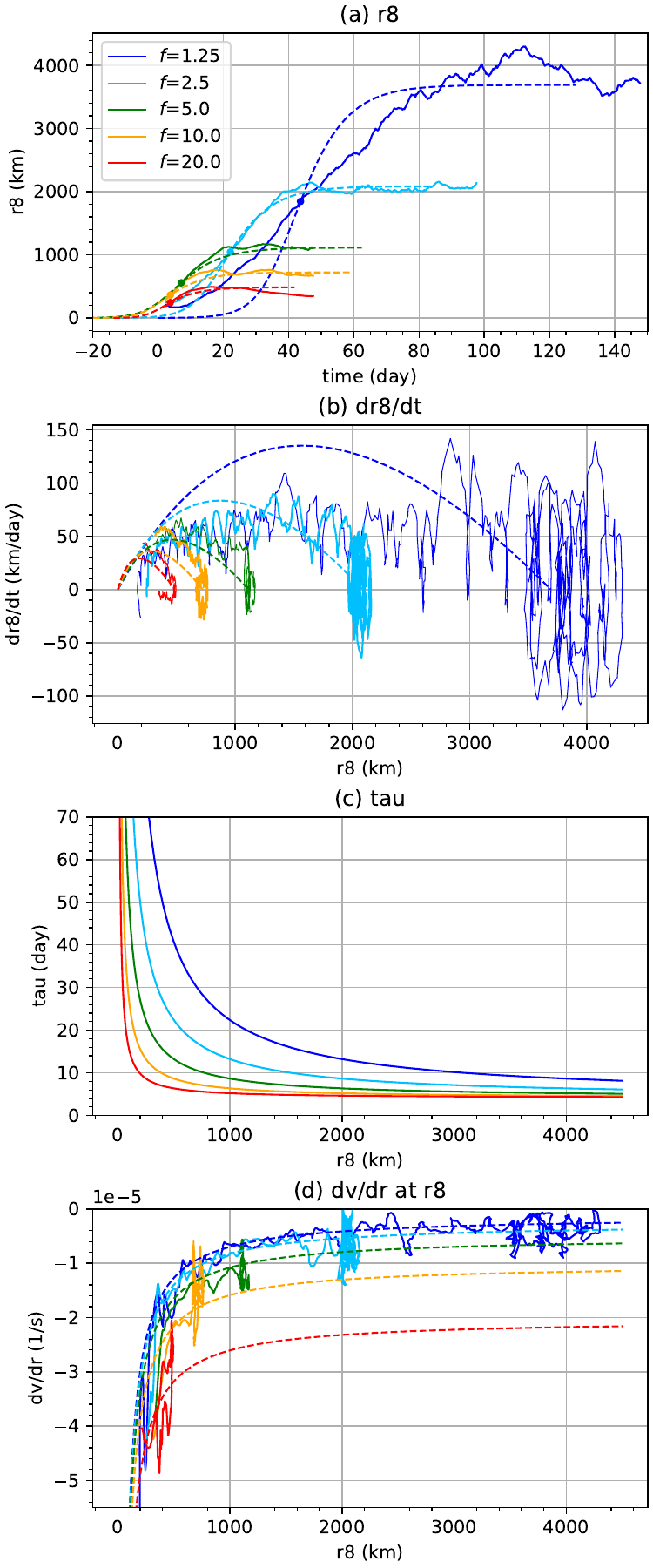}\\
 \caption{(a) Time evolution of 120-h running averaged $r_8$ (km) in FCOR (solid) and analytical expansion model prediction (dashed) taking $r_{t,eq}$ equal to equilibrium sizes of FCOR (see text). Dots mark initial condition for expansion model. (b) The corresponding expansion rate $dr_8/dt$ (km/day) in FCOR (solid, shown as the 24-h expansion rate of the 120-h running averaged $r_8$) and in expansion model (dashed). (c) $\tau_{rt}$ (day) in corresponding expansion model. (d) $\partial v/\partial r$ (1/s) at $r_8$ in FCOR (solid) and in corresponding expansion model (dashed). Warmer color means higher $f$ (see legend, $10^{-5}$ s$^{-1}$).
 }\label{fig:FCOR_size_comp}
\end{figure}

An equivalent comparison as in Fig. \ref{fig:TTPP_size_comp} is performed with FCOR (Fig. \ref{fig:FCOR_size_comp}). First, we note that equilibrium size in FCOR scales approximately with a $1/f$ scaling, with the time scale of expansion longer with lower $f$. 
Here we define equilibrium size as the average $r_8$ during 130-150 days for $f=1.25\times10^{-5}$ s$^{-1}$,  80-100 days for $f=2.5\times10^{-5}$ s$^{-1}$,   40-50 days for $f=5\times10^{-5}$ s$^{-1}$, 20-40 days for $f=10\times10^{-5}$ s$^{-1}$, and 15-20 days for $f=20\times10^{-5}$ s$^{-1}$. An earlier period is chosen for high $f$ cases to capture their peak sizes. 
These equilibrium sizes are used as $r_{t,eq}$ for the expansion model. Analogous to our analysis for TTPP in Fig. \ref{fig:TTPP_size_comp}, the expansion model only needs $\tau_{rt}$ then, which is determined by the same parameters in Sec. 3a. The lone exception is that $\tau_{rt}$ depends on $f$, which is set to the corresponding value in FCOR in each experiment.

The analytical solution of size evolution (Eq. \ref{eq:exmo_analy}) is then compared with FCOR (Fig. \ref{fig:FCOR_size_comp}a). 
Integration constants $r_{t0}$ and $t_0$ are set in the same manner as Fig. \ref{fig:TTPP_size_comp}a.
Overall, the expansion model again compares very well with the experiments in FCOR. The lone case that does not match as well is for $f=1.25\times10^{-5}$ s$^{-1}$, which expands more gradually than predicted by the theory similar to the low $T_{tpp}$ case. 
The model prediction of a higher expansion rate with lower $f$ is also consistent with FCOR (Fig. \ref{fig:FCOR_size_comp}b), except for the case with $f=1.25\times10^{-5}$ s$^{-1}$. 
Otherwise, the expansion model does reasonably well for $f=2.5\times10^{-5}$ s$^{-1}$ and $f=5\times10^{-5}$ s$^{-1}$ (and larger $f$), which are the principal latitudes ($10^{\circ}$N and $20^{\circ}$N) of TC development on Earth. 

Note that changing $f$ changes $\partial v/\partial r$ at $r_8$ as well (Fig. \ref{fig:FCOR_size_comp}d). The wind profile is steeper (larger slope) for larger $f$ (Fig. \ref{fig:exmo_wind}b), which is also observed in FCOR. 
The time scale $\tau_{rt}$ is proportional to both $(-\partial v/\partial r)$ and $1/f$ itself (Eq. \ref{eq:tau_rt}). Hence, the observation that $\tau_{rt}$ increases monotonically as $f$ decreases (Fig. \ref{fig:FCOR_size_comp}c) indicates that the effect of $1/f$ on $\tau_{rt}$ dominates that of $(-\partial v/\partial r)$. Note the difference of $\tau_{rt}$ will vanish for very large $r_t$ as expected by the property of ($-\frac{\partial v}{\partial r}$) in Appendix A. The longer expansion time scale ($\tau_{rt}$) with lower $f$ in the expansion model is also consistent with the experiments in FCOR.

Overall, the expansion model prediction compares well with the simulation experiments in Figs. \ref{fig:TTPP_size_comp}-\ref{fig:FCOR_size_comp}. We conclude that: 
\begin{enumerate}
\item Quantity $\tau_{rt}$ provides a reasonable time scale for expansion ($10\sim15$ days for 20$^\circ$N); 
\item Quantity $r_{t,eq}$ can be assumed constant with varied $V_p$ and $f$ (to a lesser extent);
\item The expansion model predicts a reasonable size evolution and expansion rate by capturing the qualitative and quantitative behavior of local spinup rate and local slope of wind profile.
\end{enumerate}
Despite the good performance, we note that for very large storms (very cold $T_{tpp}$ or very small $f$) the model tends to overestimate the simulated expansion rate, albeit for values of these parameters that are not associated with real storms on Earth.

\subsection{
Prediction of equilibrium size
}

The above subsection indicates that the expansion model works well given a known value of $r_{t,eq}$ taken from the simulations. We next test how well the model can predict $r_{t,eq}$ from environmental parameters based on the parameterization of $(M_{ew}/\rho_w)_{eq}$ (Eq. \ref{eq:fitted_Mewovrho}) in Sec. 2d. 
The equilibrium periods for $V_p$ in TTPP and FCOR are the same as in Sec. 3b. 

\begin{figure}[t]
 \noindent\includegraphics[width=0.45\textwidth,angle=0]{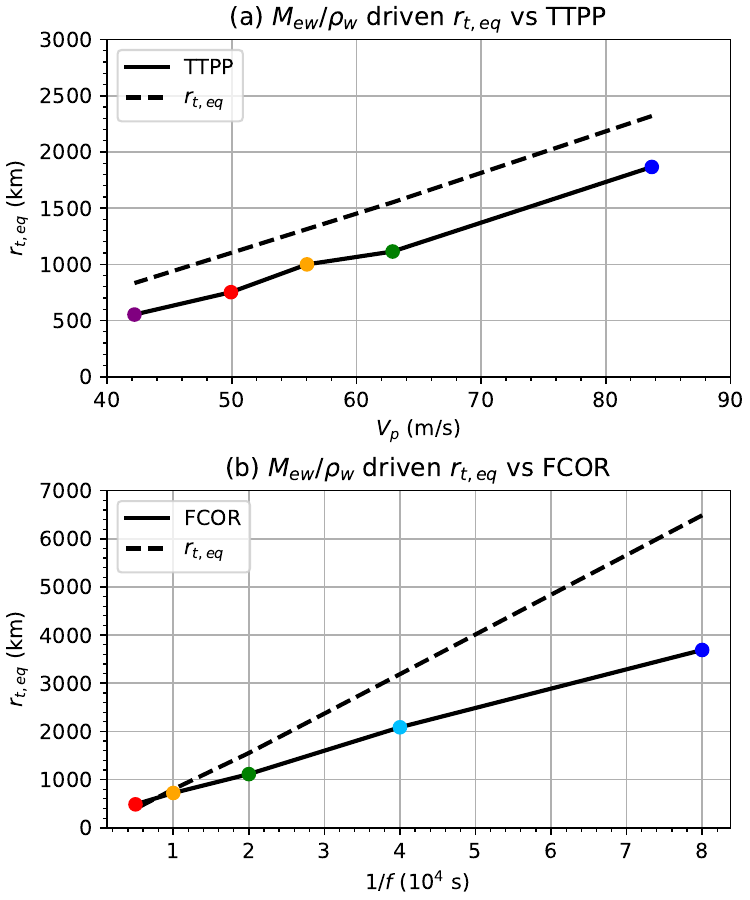}\\
 \caption{(a) Equilibrium sizes of TTPP simulations (solid) and the predicted $r_{t,eq}$  
 (dashed). (b) as in (a) but for FCOR. Dots mark different cases in TTPP and FCOR. In (a), (b) colors have the same meaning as Figs. \ref{fig:TTPP_size_comp}, \ref{fig:FCOR_size_comp}, respectively.
 }\label{fig:rteq_Mewovrho_eq_TTPP_FCOR}
\end{figure}

The resulting predictions for $r_{t,eq}$ are compared with equilibrium $r_8$ in TTPP (Fig. \ref{fig:rteq_Mewovrho_eq_TTPP_FCOR}a) and FCOR (Fig. \ref{fig:rteq_Mewovrho_eq_TTPP_FCOR}b). 
The predictions for $r_{t,eq}$ reasonably follow the simulated values in both TTPP and FCOR, with a closer match in the former. Specifically, $r_{t,eq}$ increases with $V_p$ and is proportional to $1/f$, though the latter dependence is a bit weaker than a pure linear dependence on 1/f. Thus, $r_{t,eq}$ can in principle be estimated from environmental parameters.

\subsection{Time-dependent $r_{t,eq}$}
\begin{figure}[t]
 \noindent\includegraphics[width=20pc,angle=0]{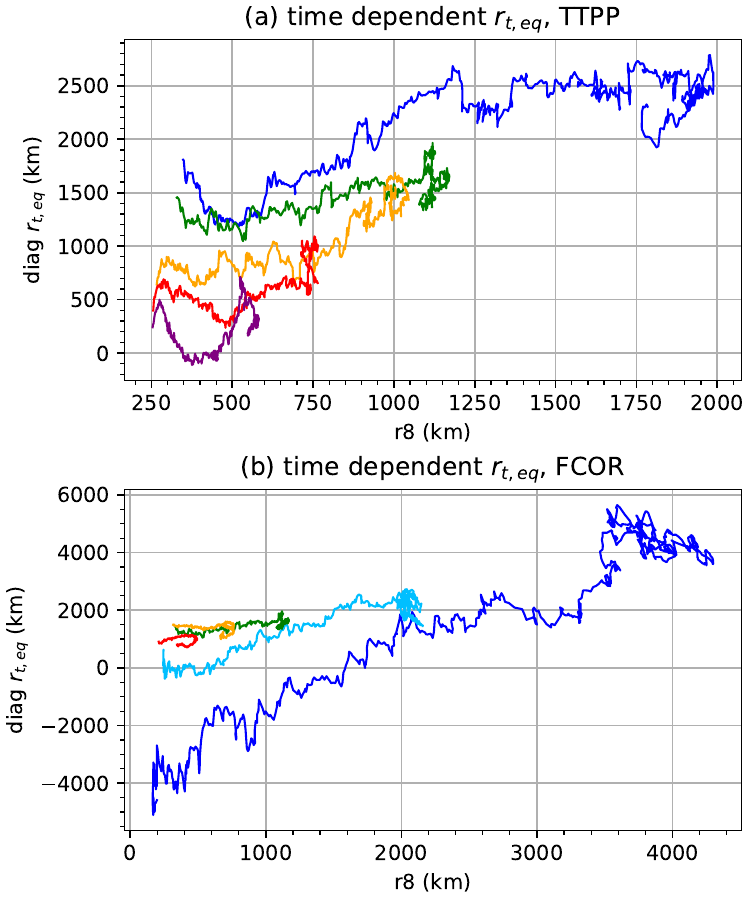}\\
 \caption{Time-dependent $r_{t,eq}$ (km) as a function of $r_8$  (km) for (a) TTPP and (b) FCOR.
 Colors in (a) and (b) have the same meaning as Figs. \ref{fig:TTPP_size_comp} and \ref{fig:FCOR_size_comp}, respectively. See text for details.
 }\label{fig:time_dependent_rteq}
\end{figure}

Though in the above analyses $r_{t,eq}$ has been treated as a time-independent constant, it can vary with time and size following Eq. (\ref{eq:rteq}b). The time-dependent $r_{t,eq}$ determined by Eq. (\ref{eq:rteq}b) using simulated $\mathcal{Q}_{lat}$ and $r_8$ and all other parameters same as in Sec. 3a (except that $f$ is set to the corresponding value in FCOR) are examined. The results are shown in Fig. \ref{fig:time_dependent_rteq}. 
Time-dependent $r_{t,eq}$ is approximately constant during expansion, consistent with the assumption of a constant $r_{t,eq}$ in expansion model, except a substantial increase in $T_{tpp}=163$ K and small $f$ (especially $f=1.25\times10^{-5}$ s$^{-1}$) cases, which explains why expansion rate is over-estimated by expansion model in these cases Figs. \ref{fig:TTPP_size_comp}b and \ref{fig:FCOR_size_comp}b. 
Notably, the diagnosed time-dependent $r_{t,eq}$ can be negative (which is obviously incorrect) and most evident for $f=1.25\times 10^{-5}$ s$^{-1}$ case. 
This simply indicates that, when $f$ is small, the expansion model misses some important process (likely eddy momentum flux in tangential wind budget) other than latent heating in favor of TC expansion while over-estimating latent heating. 
Additionally, it is reasonable that Eq. (\ref{eq:rteq}b) does not produce the exact equilibrium sizes in simulations because of the simplifications of the expansion model.

\section{Physical interpretation of the model}

\subsection{Physical meaning of $r_{t,eq}$}
Eq. (\ref{eq:rteq}a) provides an expression for equilibrium size, which is derived independently of TC PS (Wang et al. 2022) but shares many similar properties. In this subsection, we further quantitatively discuss the physical meaning of $r_{t,eq}$.

\begin{figure}[t]
 \noindent\includegraphics[width=0.45\textwidth,angle=0]{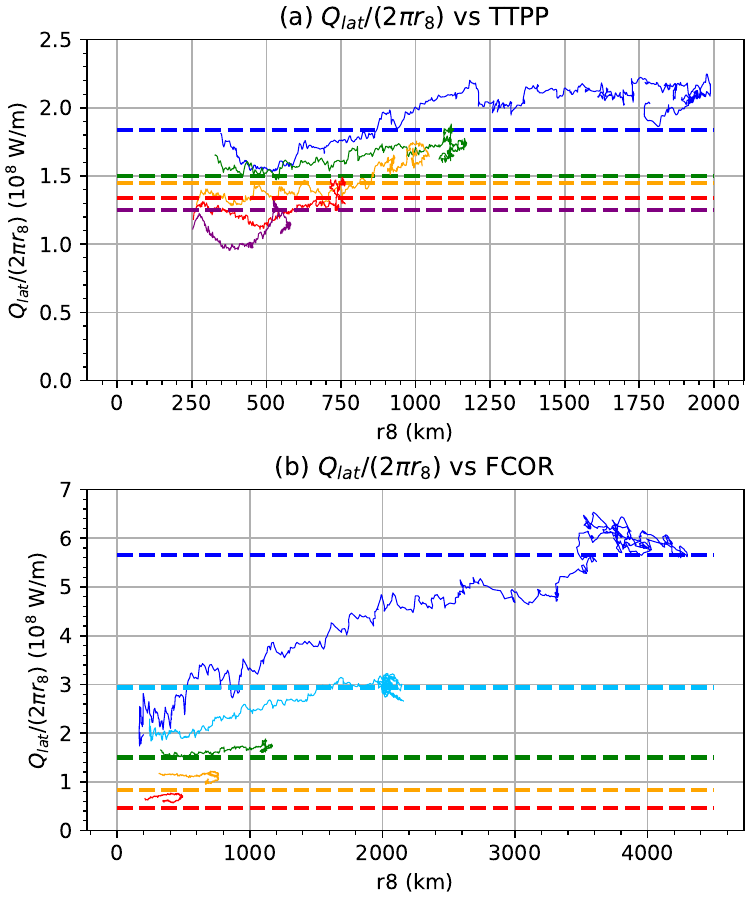}\\
 \caption{(a)  $\mathcal{Q}_{lat}/(2\pi r_8)$ ($10^8$ W/m) in TTPP (solid) and those correspondingly predicted by expansion model (dashed, see text for details). Colors have the same meaning as Fig. \ref{fig:TTPP_size_comp}. (b) same as (a), but for FCOR. Colors have the same meaning as Fig. \ref{fig:FCOR_size_comp}.  
}\label{fig:TTPP_FCOR_Qlat_ov_2pirt}
\end{figure}

In essence, $r_{t,eq}$ principally depends on (or reflected by) latent heating rate, radiative cooling rate and $f$ (Eq. \ref{eq:rteq}b)\footnote{Quantity $\mathcal{Q}_{lat}/r_t$ is completely determined by $r_{t,eq}$ through Eq. (\ref{eq:rteq}b). Thus the present model does not allow $\mathcal{Q}_{lat}$ to deviate from its ``expected" value, though it can happen in nature.}. All else equal, $r_{t,eq}$ scales directly with $\mathcal{Q}_{lat}/(2\pi r_t)$ (Eq. \ref{eq:rteq}b) and hence a larger equilibrium size should be associated with a larger $\mathcal{Q}_{lat}/(2\pi r_t)$ throughout the expansion. 
This is evident in the TTPP experiments: $\mathcal{Q}_{lat}/(2\pi r_8)$ is indeed systematically higher with lower $T_{tpp}$, corresponding to larger equilibrium sizes (Fig. \ref{fig:TTPP_FCOR_Qlat_ov_2pirt}a, solid lines). 
Conversely, if one sets $r_{t,eq}$ to an externally-defined value (as in Sec. 3b), the model predicts latent heating (Eq. \ref{eq:rteq}b) that drives the storm towards its equilibrium size with all required parameters (see Sec. 2e) identical to Sec. 3a. 
The expansion model predictions of $\mathcal{Q}_{lat}/(2\pi r_8)$ (Fig. \ref{fig:TTPP_FCOR_Qlat_ov_2pirt}a, dashed lines) captures the systematic variation found in TTPP, demonstrating that larger latent heating rate during the expansion leads to larger equilibrium size. For TTPP, the enhanced latent heating rate arises principally because of the enhanced overturning mass flux (Eq. \ref{eq:Qlat}), which is larger at higher potential intensity (Eq. \ref{eq:fitted_Mewovrho}).
And this also leads to a higher expansion rate of TCs with lower $T_{tpp}$, consistent with the analysis in Sec. 3c.

Similar dependencies are evident when $f$ is varied. Since $r_{t,eq}$ scales as $1/f$ (Eqs. \ref{eq:r0_eq},\ref{eq:rteq_solved},\ref{eq:fitted_Mewovrho}), $\mathcal{Q}_{lat}/(2\pi r_t)$ should also scale with $1/f$ (Eq. \ref{eq:rteq}b). 
This is found qualitatively true in both FCOR (Fig. \ref{fig:TTPP_FCOR_Qlat_ov_2pirt}b, solid lines) and in expansion model prediction (Fig. \ref{fig:TTPP_FCOR_Qlat_ov_2pirt}b, dashed lines) with $r_{t,eq}$ set to the equilibrium size of FCOR as in Sec. 3b, $f$ set to the corresponding value in FCOR, and all other parameters identical to Sec. 3a. The $f=1.25\times10^{-5}$ s$^{-1}$ simulation case deviates more strongly in that this quantity is substantially smaller than the model predicted value during the main expansion stage when $f=1.25\times10^{-5}$ s$^{-1}$; this is consistent with its deviation from the model prediction of size itself discussed in Sec. 3d.



To summarize, in the expansion model, equilibrium size is effectively modulated by $\mathcal{Q}_{lat}/(2\pi r_t)$, and the expansion rate is modulated by the equilibrium size. 
For a given initial size, a larger $\mathcal{Q}_{lat}/(2\pi r_t)$ translates to a larger $r_{t,eq}$ and thus a larger expansion rate. In practice, if a TC moves to a more favorable environment for convection, then its expansion rate would be expected to increase and its expected equilibrium size also increases, as quantitatively described by Eq. (\ref{eq:rteq}b). This interpretation is consistent with the behavior of observed storms, which have been found to expand when convection is enhanced outside of the storm inner core across a variety of distinct forcing mechanisms (Maclay et al. 2008\nocite{Maclay_DeMaria_VonderHaar_2008}).

The dependence of $r_{t,eq}$ on the latent heating rate ($\mathcal{Q}_{lat}/(2\pi r_t)$) is complementary to TC PS model, with the former being a reflection of the volume (or mass) integrated processes of the system instead of a single parcel's cycle.
As noted above, the larger latent heating rate produces a larger $r_{t,eq}$ (with the same $f$) as it is associated with a larger $V_p$, consistent with $V_{Carnot}/f$ scaling (Wang et al. 2022). 
Meanwhile, the effect of $C_d$ in the size scaling is substantially reduced because an increase of $C_d$ reduces the storm intensity but increases the inflow angle and inflow depth under the eyewall at the same time, such that $M_{ew}$ is remains relatively constant.
\subsection{Expansion mechanism}

The model (Eq. \ref{eq:exmodel_general}) physically assumes that TC size expansion is driven principally by latent heating, which drives low-level lateral inflow that imports absolute vorticity to expand the storm. In this section, we test how the model predicts the dependencies of inflow velocity $u_t$, local spinup rate $\partial v/\partial t$ and expansion rate $dr_t/dt$ on latent heating rate, and how they compare with simulations.

While the final version of the model is predictive based on environmental parameters alone, the model implicitly provides a quantitative dependence of a response of expansion rate to a change of latent heating $\mathcal{Q}_{lat}$. This is useful when a TC experiences an inner-core structural variation such as secondary eyewall formation (e.g. Kossin and Sitkowski 2009\nocite{Kossin_Sitkowski2009}). From Eq. (\ref{eq:ut_energetics}a), we have, holding $r_t$ constant,
\begin{subequations}\label{eq:sens_expa_Qlat}
    \begin{align}
        \frac{\partial u_t}{\partial [\mathcal{Q}_{lat}/(\pi r_t^2)]}=-\frac{r_t}{2 h_w\rho_iT_{e,lat}\Delta s_d}\\
        \frac{\partial (\partial v_t/\partial t)}{\partial [\mathcal{Q}_{lat}/(\pi r_t^2)]}=-f\frac{\partial u_t}{\partial [\mathcal{Q}_{lat}/(\pi r_t^2)]}\\
        \frac{\partial (dr_t/dt)}{\partial [\mathcal{Q}_{lat}/(\pi r_t^2)]}=\frac{-f}{(-\frac{\partial v}{\partial r})}\frac{\partial u_t}{\partial [\mathcal{Q}_{lat}/(\pi r_t^2)]}
    \end{align}
\end{subequations}
where we have expressed the change of latent heating as that per unit area. Eq. (\ref{eq:sens_expa_Qlat}) states that, all else being equal, an increase of latent heating leads to an increase of lateral inflow magnitude and thus expansion rate. 
The derivatives in Eq. (\ref{eq:sens_expa_Qlat}) only depends on $r_t$ and not $\mathcal{Q}_{lat}$ itself. These relations can be directly compared to numerical simulations.


For $u_t$, Eq. (\ref{eq:sens_expa_Qlat}a) indicates that the derivative $\frac{\partial u_t}{\partial [\mathcal{Q}_{lat}/(\pi r_t^2)]}$ is proportional to $r_t$, meaning the sensitivity increases with $r_t$. $\frac{\partial (\partial v/\partial t)}{\partial [\mathcal{Q}_{lat}/(\pi r_t^2)]}$ is also proportional to $r_t$ (Eq.~\ref{eq:sens_expa_Qlat});  $\frac{\partial (dr_t/dt)}{\partial [\mathcal{Q}_{lat}/(\pi r_t^2)]}$ is further modified by $(-\frac{\partial v}{\partial r})^{-1}$. As $(-\frac{\partial v}{\partial r})^{-1}$ also increases with $r_t$, then $\frac{\partial (dr_t/dt)}{\partial [\mathcal{Q}_{lat}/(\pi r_t^2)]}$ also increases with $r_t$. As these derivatives are all independent of $\mathcal{Q}_{lat}$, then $u_t$, $\partial v/\partial t$ and $dr_t/dt$ will be linear functions of $\mathcal{Q}_{lat}/(\pi r_t^2)$, for any fixed $r_t$. 

\begin{figure}[t]

\noindent\includegraphics[width=30pc,angle=0]{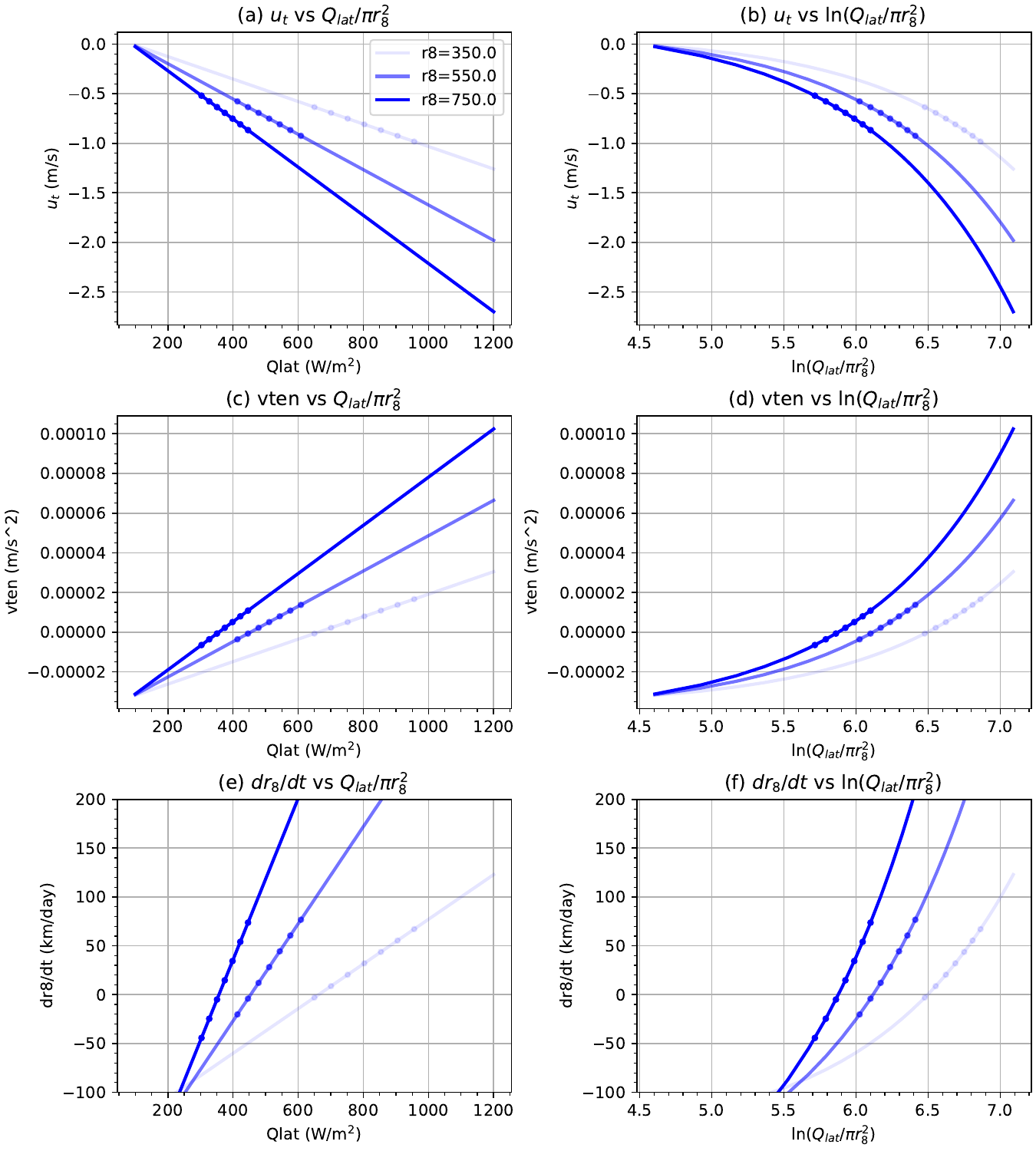}\\
 \caption{Prediction from expansion model with $f=5\times10^{-5}$ s$^{-1}$ of the relationship between latent heating rate $\mathcal{Q}_{lat}/(\pi r_8^2)$ and (a) $u_t$ (m/s) at $r_8$, (c) $\partial v/\partial t$ (m/s$^2$) at $r_8$, (e) $dr_8/dt$ (km/day). (b), (d), (f) are equivalent to (a), (c), (e) respectively, but taking $\ln(\mathcal{Q}_{lat}/(\pi r_8^2))$ along the x-axis. Results for three fixed $r_t$ values are shown: 350, 550, 750 km. Dots mark the prediction in Figs. \ref{fig:model_ideal_1}-\ref{fig:model_ideal_2}.
 }\label{fig:sens_Qlat_exmo}
\end{figure}

These ideal relations are shown in Fig. \ref{fig:sens_Qlat_exmo} with $f=5\times10^{-5}$ s$^{-1}$ and other environmental parameters identical to Sec. 3a, for three representative values of $r_t=r_8$: 350, 550 and 750 km. As discussed above, $\frac{\partial u_t}{\partial [\mathcal{Q}_{lat}/(\pi r_t^2)]}$, $\frac{\partial (\partial v/\partial t)}{\partial [\mathcal{Q}_{lat}/(\pi r_t^2)]}$, and  $\frac{\partial (dr_t/dt)}{\partial [\mathcal{Q}_{lat}/(\pi r_t^2)]}$ are all larger in magnitude when the TC is larger (slopes in Fig. \ref{fig:sens_Qlat_exmo}a,c,e). 
Specifically, it is of practical interest how $u_t$, $\partial v/\partial t$ and $dr_8/dt$ respond to a proportional change of latent heating. 
This is equivalent to seeking $\frac{\partial u_t}{\partial \ln[\mathcal{Q}_{lat}/(\pi r_t^2)]}$, $\frac{\partial (\partial v/\partial t)}{\partial \ln[\mathcal{Q}_{lat}/(\pi r_t^2)]}$, and $\frac{\partial (dr_t/dt)}{\partial \ln[\mathcal{Q}_{lat}/(\pi r_t^2)]}$ at constant $r_t$ (the slope of the curves in Fig. \ref{fig:sens_Qlat_exmo}b,d,f). It is evident that $u_t$, $\partial v/\partial t$ and $dr_8/dt$ becomes more sensitive to a proportional change of latent heating rate when TC is larger, indicating that size changes of large TCs are more variable. Quantitatively, from Fig. (\ref{fig:sens_Qlat_exmo}f), a 10\% proportional change of latent heating rate ($\ln(\mathcal{Q}_{lat}/(\pi r_8^2))$ change by 0.1) would induce a change in $dr_8/dt$ of tens (about 10-30) of kilometers per day.

\clearpage

\clearpage

\begin{figure}[t]

\noindent\includegraphics[width=30pc,angle=0]{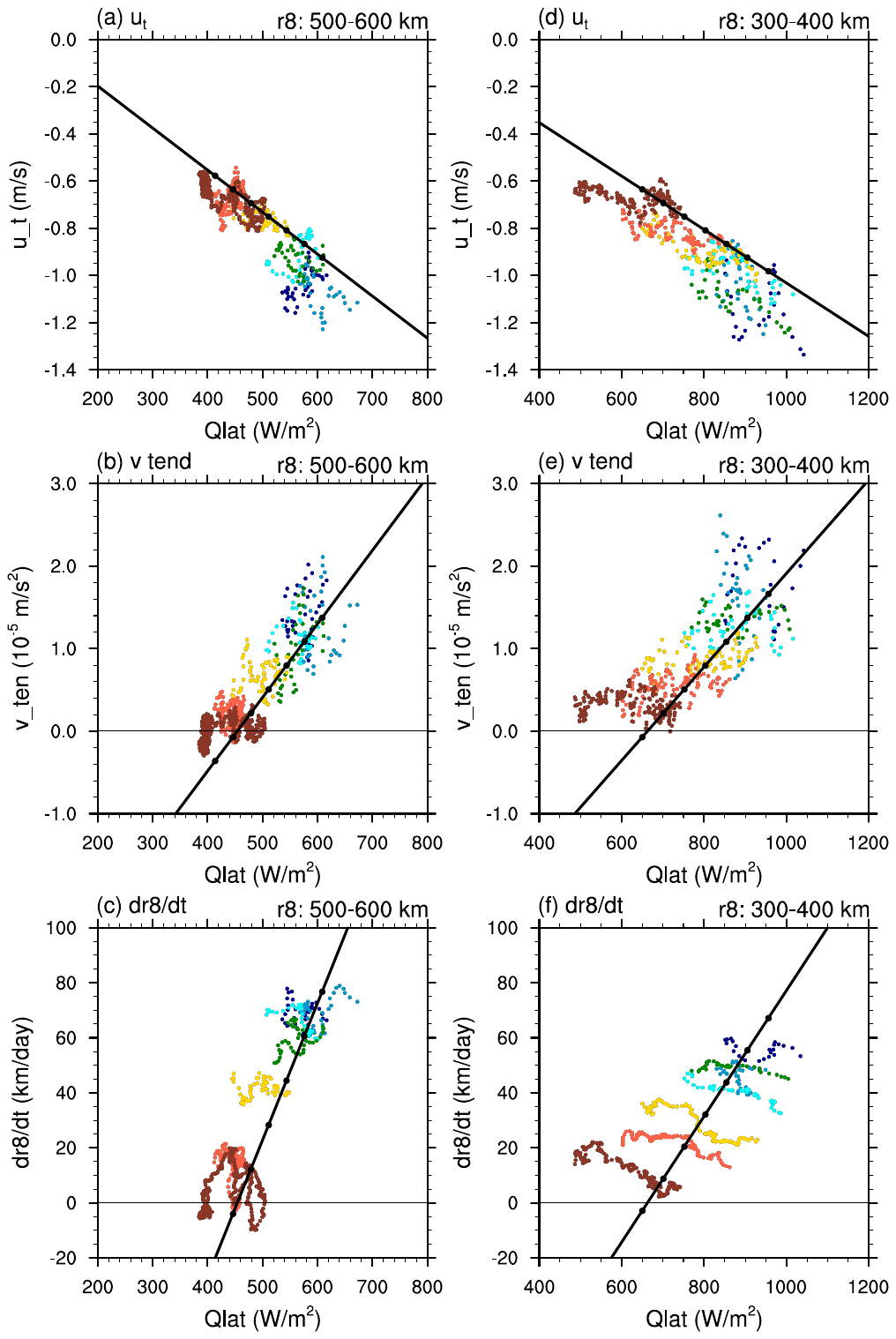}\\
 \caption{(a) Colored dots shows $u_t$ (m/s) and $\mathcal{Q}_{lat}/(\pi r_8^2)$ (W/m$^2$) when $r_8$ above 500 km and below 600 km in TTPP; warmer color means higher $T_{tpp}$. Black line shows expansion model prediction with $r_8=550$ km and black dots mark cases in Figs. \ref{fig:model_ideal_1}-\ref{fig:model_ideal_2}. (b) as in (a), but for $\partial v/\partial t$ (m/s$^2$) at $r_8$. (c) as in (a) but for $dr_8/dt$ (km/day), shown as the 24-h expansion rate of the 120-h running averaged $r_8$. (d)-(f), as in (a)-(c), but with $r_8=350$ km in expansion model and $r_8$ from 300 to 400 km in TTPP. Note all cases in TTPP are shown.
 }\label{fig:sens_Qlat_TTPP_56_34}
\end{figure}

The linear relations are compared with TTPP, shown in Fig. \ref{fig:sens_Qlat_TTPP_56_34} . The $r_8$ is first taken as 550 km and data in TTPP is collected with $r_8$ from 500 to 600 km (Fig. \ref{fig:sens_Qlat_TTPP_56_34}a-c). 
This size corresponds to relatively large (above median $\sim400$ km, Schenkel et al. 2023) TCs on earth. We see an overall nice match of both the slope (sensitivity) and the absolute value between expansion model prediction and TTPP for all of $u_t$, $\partial v/\partial t$ and $dr_8/dt$. TCs with lower $T_{tpp}$ in TTPP are associated with larger latent heating rate, which leads to stronger inflow velocity, local spinup rate and expansion rate, consistent with the evolution of $r_8$ in Fig. \ref{fig:TTPP_size_comp}. This also supports the overall hypothesis of the expansion model that latent heating drives expansion. Note the expansion rate is rather sensitive to latent heating.  For $\mathcal{Q}_{lat}/(\pi r_8^2)=500$ W/m$^2$ in Fig. \ref{fig:sens_Qlat_TTPP_56_34}c, a 20\% change of latent heating may either double the expansion rate or terminate expansion.

\clearpage

\clearpage

The same analysis is repeated but with $r_8=350$ km in the expansion model and $r_8$ from 300 to 400 km in TTPP (Fig. \ref{fig:sens_Qlat_TTPP_56_34}d-f). This size corresponds to relatively small (below median) TCs on earth. The predictions of $u_t$ and $\partial v/\partial t$ as linear functions of $\mathcal{Q}_{lat}/(\pi r_8^2)$ are still valid; The discrepancy for $dr_8/dt$ appears larger compared to $r_8=550$ km (Fig. \ref{fig:sens_Qlat_TTPP_56_34}a-c). 
The value of $dr_8/dt$ is less sensitive to latent heating in TTPP than in the expansion model prediction. This likely results from an underestimation of the magnitude of $\partial v/\partial r$ by the model (Fig. \ref{fig:TTPP_size_comp}c).



\begin{figure}[t]

\noindent\includegraphics[width=20pc,angle=0]{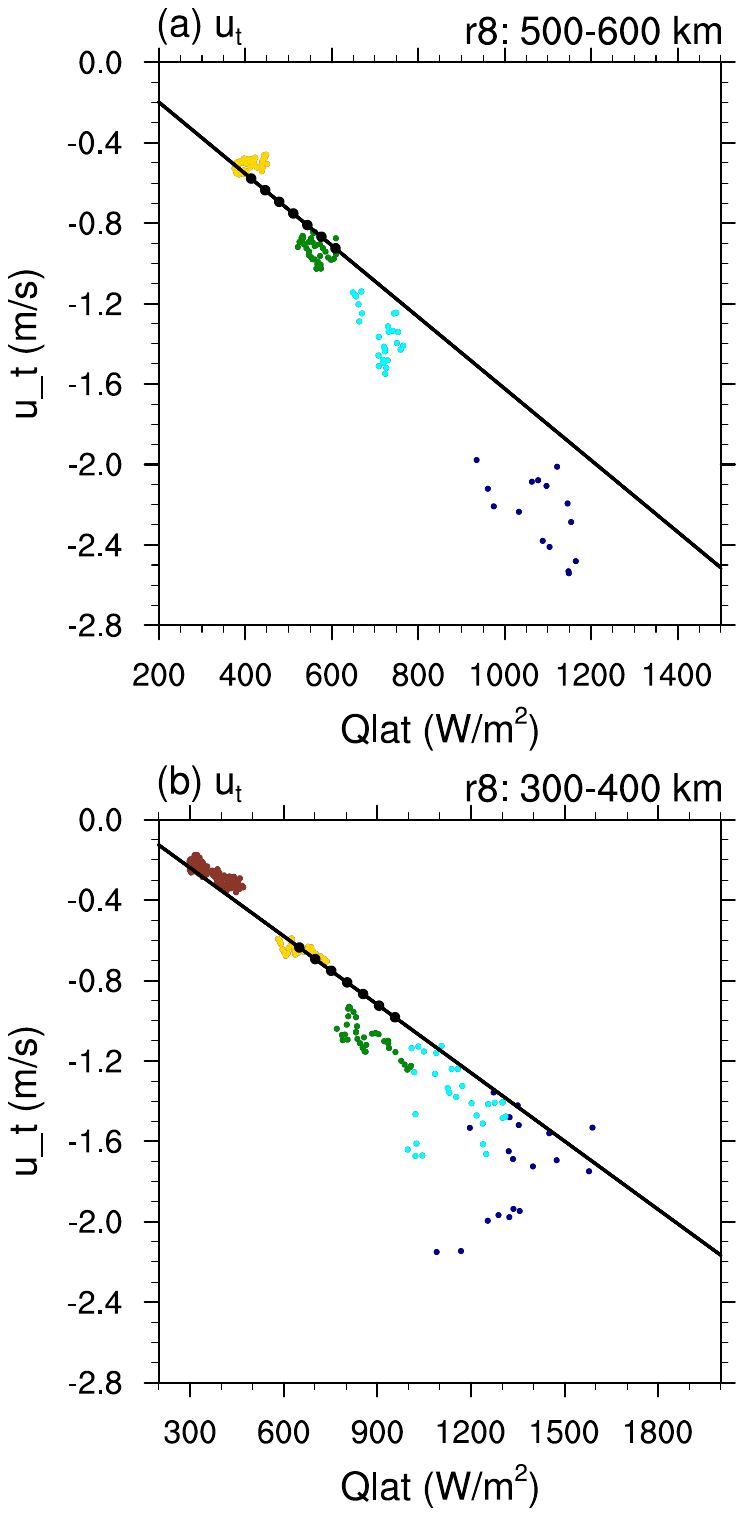}\\
 \caption{(a) Colored dots shows $u_t$ (m/s) and $\mathcal{Q}_{lat}/(\pi r_8^2)$ (W/m$^2$) when $r_8$ above 500 km and below 600 km in FCOR; warmer color means higher $f$. Black line shows expansion model prediction with $r_8=550$ km and black dots mark cases in Figs. \ref{fig:model_ideal_1}-\ref{fig:model_ideal_2}. (b) as in (a), but with $r_8$ above 300 km and below 400 km for FCOR and $r_8=350$ km for expansion model.
 }\label{fig:sens_Qlat_FCOR}
\end{figure}

The dependence of $u_t$ on latent heating rate can also be tested against FCOR, as (also at constant $r_t$) $\frac{\partial u_t}{\partial [\mathcal{Q}_{lat}/(\pi r_t^2)]}$ does not depend on $f$ (but $\frac{\partial (\partial v/\partial t)}{\partial [\mathcal{Q}_{lat}/(\pi r_t^2)]}$ and $\frac{dr_8/dt}{\partial [\mathcal{Q}_{lat}/(\pi r_t^2)]}$ depend on $f$). 
Analogous to the comparison for TTPP in Fig. \ref{fig:sens_Qlat_TTPP_56_34}, the comparison is shown in 
Fig. \ref{fig:sens_Qlat_FCOR}. The expansion model matches the simulations in both the absolute value and the slope of the relations.  Notably, the latent heating rate is substantially larger with smaller $f$, consistent with Fig. \ref{fig:TTPP_FCOR_Qlat_ov_2pirt}b. 

Overall, then, the expansion model can predict the dependence of $u_t$, $\partial v/\partial t$ and $dr_8/dt$ on latent heating rate, especially for $u_t$. This provides experimental evidence that lateral inflow velocity and its resulting spinup is indeed driven principally by latent heating in the TC, with their quantitative dependence described by Eq. (\ref{eq:ut_energetics}).

\subsection{Model sensitivity to $\Delta s_d$}

In this subsection, we test the sensitivity of the expansion model (Sec. 2) to $\Delta s_d$, an important parameter that modulates both $r_{t,eq}$ (Eq. \ref{eq:rteq_solved}) and $\tau_{rt}$ (Eq. \ref{eq:tau_rt}) through $B$. 
Increasing $\Delta s_d$ decreases $B$, which decreases $r_{t,eq}$ and increases $\tau_{rt}$, which both cause the expansion rate to decrease (Eq. \ref{eq:exmodel_general}).  

Physically, 
the overturning acts to export entropy because inflow brings in lower entropy air while outflow takes out higher entropy air. This depends on $\Delta s_d$.  If $\Delta s_d$ is larger, then the overturning circulation can be less intense to achieve the same net export, which means smaller magnitude of inflow velocity and hence slower expansion.
The expansion model works reasonably partly because in a mature TC a main portion of the inflow mass flux is confined to low levels. Hence, overturning circulations are very efficient at exporting entropy; this behavior can be characterized as having a strongly positive (i.e. stable) Gross Moist Stability (Raymond et al. 2009\nocite{Raymond_etal_2009}). 

To perform the sensitivity test, we vary $\Delta s_d$ about the base value (187.1 J/K/kg) in Sec. 3 by multiplying it by 0.5, 0.75, 1.0, 1.25, 1.5. In the calculation, $\partial v/\partial r$ is not modified (and hence $\xi_0$ remains fixed to the value in Sec. 3, though $A$, $B$ and $\xi$ vary with $\Delta s_d$) for simplicity. 
Quantity $r_{t,eq}$ is determined by the $(M_{ew}/\rho_w)_{eq}$ (Eq. \ref{eq:fitted_Mewovrho}) in the base environment through Eqs. (\ref{eq:r0_eq}), (\ref{eq:rteq_solved}).

\begin{figure}[t]
 \noindent\includegraphics[width=0.45\textwidth,angle=0]{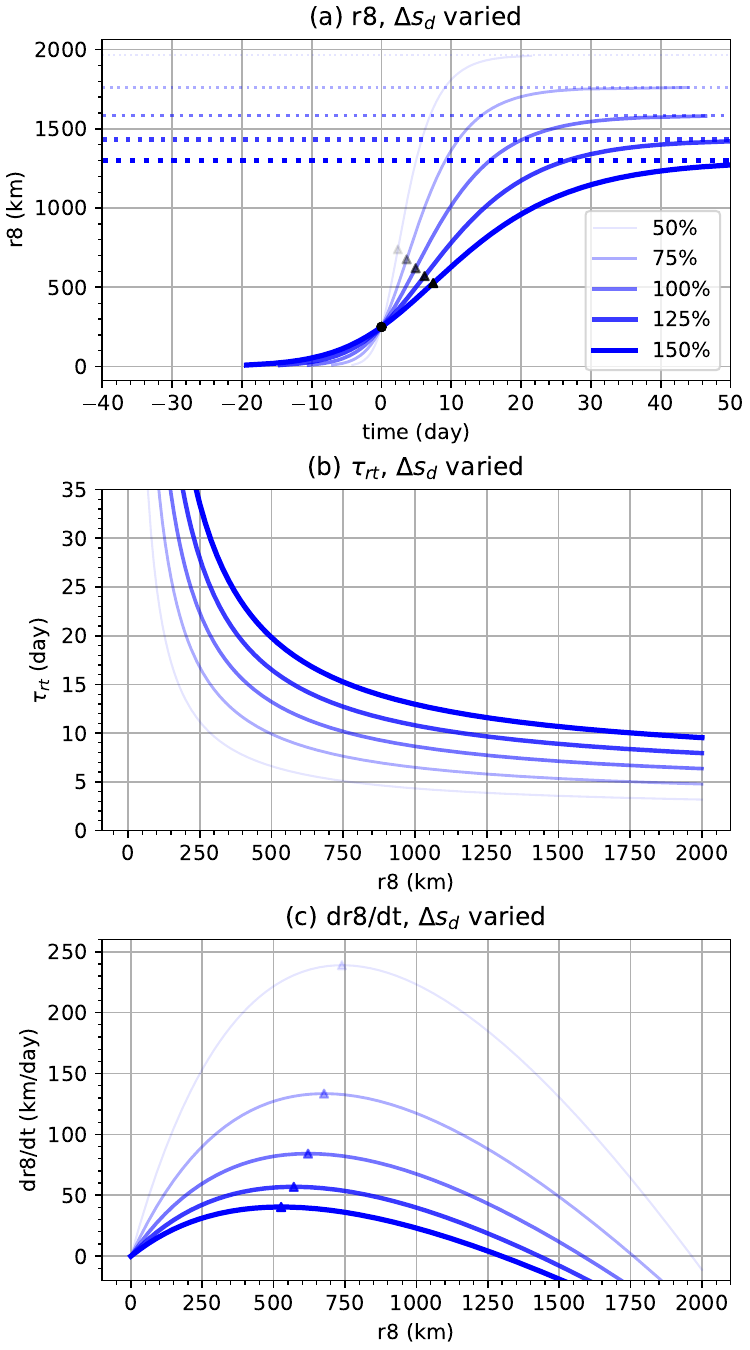}\\
 \caption{Same as Fig. \ref{fig:model_ideal_1} but with $\Delta s_d$ varied from 50\% to 150\% of its base value with an inverval of 25\%, see text for details. In (a)-(c), thicker and more opaque lines mark higher values of $\Delta s_d$. 
 }\label{fig:model_ideal_vary_delta_sd}
\end{figure}

The results are shown in Fig. \ref{fig:model_ideal_vary_delta_sd}. When $\Delta s_d$ varies from 50\% to 150\% of its base value, $r_{t,eq}$ decreases from 2000 to 1300 km; quantity $\tau_{rt}$ increases by a factor of 3. Expansion rate decreases from over 200 km/day to below 50 km/day. 
Though the overall values of expansion rate, $r_{t,eq}$ and $\tau_{rt}$ are still reasonable, it is evident that the expansion process can be directly modulated by $\Delta s_d$. This implies that in a warmer climate, a likely increase of $\Delta s_d$ would partially offset the effect of an increase of latent heating rate that drives faster expansion. Such questions are an important avenue of future work.

\section{Summary and discussion}

In this paper, a predictive analytic model for tropical cyclone size expansion on the $f$-plane is proposed and its overall behavior is tested against numerical simulations varying tropopause temperature (TTPP) and Coriolis parameter (FCOR). The expansion rate is described by a simple kinematic relation and is equal to the ratio of local spinup rate of tangential wind ($\partial v/\partial t$) at outer radius $r_t$ to the negative slope of the wind profile at that radius. 
The model predicts that size expansion is driven by latent heating (dominated by the eyewall) and suppressed by radiative cooling. This prediction is achieved by combining the tangential velocity budget at $r_t$ with the volume-integrated entropy (heat) budget inside of $r_t$, with the two linked via a simple relationship between outer storm size and the upward mass flux of the overturning circulation. 
Area-integrated latent heating is proportional to $r_t$, while area-integrated radiative cooling is proportional to $r_t^2$, such that the storm size eventually reaches an equilibrium ($r_{t,eq}$). The size expansion rate is the ratio of the difference between $r_{t,eq}$ and present size to a time scale $\tau_{rt}$, and both parameters may be defined from environmental parameters alone. 
Key takeaways are as follows:
\begin{itemize}
    \item The model yields a predictive, analytic solution for the evolution of storm size towards its equilibrium size (both larger or smaller than present size) given environmental parameters and an externally set $r_{t,eq}$. This solution performs well in predicting the simulated size evolutions and expansion rates in simulations across a range of values of tropopause temperature (and hence potential intensity) and Coriolis parameter. 
    \item The model successfully produces a characteristic expansion rate for $r_8$ of tens of kilometers per day with reasonable environmental parameters (Fig. \ref{fig:model_ideal_1}), in line with past work using data for historical storms.
    \item The model predicts that local spinup rate decreases quasi-linearly with expansion, and that  $\frac{\partial v}{\partial r}$ at $r_8$ decreases in magnitude with expansion, both consistent with simulations. The results taken together explain why expansion rate peaks during the early-to-mid stages of expansion rather than at the beginning, as seen in simulations. 
    \item  The model predicts higher expansion rate with higher $r_{t,eq}$ (especially when $V_p$ changes) due to the corresponding higher latent heating rate, consistent with simulations. 
    \item The model predicts reasonable time scale $\tau_{rt}$ of $10\sim15$ days when $f=5\times10^{-5}$ s$^{-1}$ and radiative cooling rate 1 K/day. The time scale is constant when varying $T_{tpp}$ and larger when $f$ is smaller, consistent with simulations.
    \item The model predicts that equilibrium size $r_{t,eq}$ increases with $\sqrt{w_{cool}}C_d^{0.47}V_p/f$, which is a scaling for the square root of equilibrium eyewall updraft mass flux and $r_{0,eq}$.
    The quantity $r_{t,eq}$ is directly linked to the latent heating rate within the TC volume and is complementary to the TC Potential Size of Wang et al. (2022) while providing clearer mechanistic insight into the process of expansion.
    \item The model predicts that enhanced latent heating will cause the storm to expand faster, consistent with past observational work finding that storms tends to expand significantly after convection is forced outside of the eyewall regardless of the forcing mechanisms (Maclay et al. 2008).
    \item The model provides mechanistic understanding for why TC size expands towards an equilibrium in the first place: the overturning circulation exports excess latent heating in the TC when it is sufficiently small, and the resulting induced low-level inflow imports absolute vorticity into the volume in excess of that lost to surface friction that spins up the tangential wind in the outer circulation. As a result, the TC expands.
\end{itemize}


Here we have taken a number of parameters as constant across our tests for simplicity to arrive at an analytically-tractable model that appears to capture the first-order behavior of the dynamics of TC size. In reality such parameters may not be perfectly constant (both in time and across experiments), a topic that could be more carefully examined within the simulations presented in this paper. Here we have evaluated this assumption for a few key parameters, but additional detailed tests of the model assumptions as well as quantifying model parameters in simulations could be tackled in a future study.

This model offers a valuable foundation for better understanding and predicting changes in storm size on Earth. For example, it is known that TCs will shrink in the absence of convection with nonzero $\beta$ (meridional gradient of $f$) due to radiation of planetary Rossby waves (Chavas and Reed 2019\nocite{Chavas_Reed_2019}; Lu and Chavas 2022\nocite{Lu_Chavas_2022}). How this shrinking effect alters the evolution of TC size on the $f$-plane is an important question for predicting changes in TC size in nature. 
Incorporation of the $\beta$ effect in the expansion model is a valuable avenue for future work.

One key assumption of the present model is the existence of an equilibrium TC outer size $r_{t,eq}$ that depends on environmental parameters on the $f$-plane.  Though a short-term equilibrium of size does exist in our simulations, a long-term existence/maintenance of equilibrium size is not conclusive in literature: Persing et al. (2019\nocite{Persing_etal_2019}) pointed out that long-term (longer than tens of days) maintenance in limited domains may require artificial source of relative angular momentum. However, in the present expansion model, $r_{t,eq}$ is valid and well-defined at any particular instance and does not require that this radius remain constant for all time (it is able to maintain for about 10 days or potentially longer in our simulations). Nonetheless, an understanding of $r_{t,eq}$ in the context of long-term TC maintenance in limited domains remains an open question. 

One open mechanistic question that we did not analyze in detail is the connection between potential intensity and updraft mass flux. In our simulations, expansion rate increases with $V_p$ (by lowering $T_{tpp}$) consistent with our theory, and the majority of the condensational heating within $r_8$ does occur within the eyewall(Appendix D). Hence, it is likely that $V_p$ modulates the strength of overturning circulation in the eyewall by modulating TC intensity, which in turns modulates boundary layer frictional convergence. However, we do not explicitly quantify the eyewall dynamics in the present model.\footnote{Thus we do expect $V_p$ to represent a characteristic intensity of the TC (see Supplemental Material), though the exact relation between $V_p$ and maximum wind speed is not considered here (see discussion of superintensity in Persing and Montgomery 2003; Li et al. 2020\nocite{Persing_Montgomery_2003,Li_Wang_etal_2020}).}. 
This also simplifies the problem a bit because potential intensity theory exists to describe eyewall structure and successfully predicts a characteristic maximum wind speed, whereas rainband activity is much less well understood and is much harder to predict. Correspondingly, although rainband activity have been reported to effectively modulate TC size expansion, it is not actively predicted in the present model and is only parameterized by a constant $\alpha_p$. 

Finally, we note that the proposed expansion model predicts that the expansion rate is a function of present size (Eq. \ref{eq:drtdt}), and hence it is independent of the preceding history of storm size including initial vortex size. This is supported by an additional set of experiments varying initial vortex size and intensity (see Supplemental Material), suggesting that $V_p$ may be a more effective factor than initial vortex structure to modulate size expansion rate.

%
%

%

\clearpage
\acknowledgments
This work is supported by National Science Foundation (NSF) AGS grant 1945113.

%
%
\datastatement

Description of CM1 model is available at https://www2.mmm.ucar.edu/people/bryan/cm1/. The specific CM1v19.2 model code with the noted modifications and the namelist for CTL simulation used in this study have been uploaded to figshare with DOI: 10.6084/m9.figshare.22674361.

%






%



\appendix[A] 
\appendixtitle{Further discussion on $\partial v/\partial r$ (Eq. \ref{eq:dvdr_solved2})}

Eq. (\ref{eq:dvdr_solved2}) indicates that $\partial v/\partial r$ is negative definite and it tends to $-\infty$ when $r_t$ tends to 0; $\partial v/\partial r$ tends to $-\frac{1}{\sigma\xi_0 v_t}$ when $r_t$ tends to $+\infty$. Specifically, Eq. (\ref{eq:dvdr_solved2}) gives:
\begin{equation}\label{eq:ddxi_dvdr}
    \frac{\partial}{\partial \xi_0}(\frac{\partial v}{\partial r}\bigg|_{v=v_t})=\frac{1}{\sigma\xi_0^2v_t}>0,
\end{equation}
which indicates that magnitude of $\partial v/\partial r$ decreases with increased $\xi_0$ and vice versa. Physically, when $C_d$ is increased or $f$ is decreased, the magnitude of $\partial v/\partial r$ will decrease. 
Additionally, when $2r_{t}\gg\xi_0 v_t^2$, then  $(\frac{\partial v}{\partial r})^{-1}|_{v=v_t}\approx-\frac{2r_tv_t\sigma\xi_0}{2r_t}=-v_t\sigma\xi_0$, giving a proportional dependence of $\partial v/\partial r$ on $f/C_d$.
When $2r_{t}\ll\xi_0 v_t^2$, then  $(\frac{\partial v}{\partial r})^{-1}|_{v=v_t}\approx-\frac{2r_t\sigma}{v_t}$, proportional to $r_t$ and independent of $\xi_0$.

\appendix[B] 
\appendixtitle{E04 model}

The E04 model provides the near-surface wind profile through a slab boundary layer model in the subsidence region where net vertical velocity is typically negative and is expressed as:
\begin{equation}\label{eq:E04}
    \frac{\partial v}{\partial r}=\frac{2 C_drv^2}{w_{cool}(r_0^2-r^2)}-f-\frac{v}{r}
\end{equation}
where $w_{cool}$ a constant clear-air subsidence velocity (positive downward) induced by radiative cooling, here set to 0.003 m/s. In the model $\frac{\partial v}{\partial r}$ is determined locally by inflow mass flux and friction and the inflow mass flux is determined by accumulated subsidence mass flux outward.

\appendix[C] 
\appendixtitle{Experimental design and processing}

Numerical experiments are performed with Cloud Model 1 (CM1; Bryan and Fritsch 2002\nocite{Bryan_Fritsch_2002}), which is a nonhydrostatic model mainly designed for idealized simulations. The model configuration is the essentially the same as Wang et al. (2022). 
The radiation is represented by applying a constant cooling rate $Q_{cool}$ of potential temperature $\theta$ where temperature is above a prescribed tropopause temperature $T_{tpp}$. Where temperature is lower than $T_{tpp}$, the $\theta$ is relaxed to the value corresponding to $T_{tpp}$ with a time scale $\tau=12$ h. A surface gustness $u_{sfc}=5$ m/s is added to 10-m wind in aerodynamic formula for surface drag and enthalpy fluxes. Morrison double-moment microphysics scheme (Morrison et al. 2005\nocite{Morrison_etal_2005}) is used.

We set environment in CTL simulation as $T_{tpp}=200$ K, $T_{SST}=300$ K, $f=5\times10^{-5}$ s$^{-1}$, $C_d=C_k=0.0015$, $Q_{cool}=1$ K/day.
In the first set of experiment (TTPP, Table \ref{t1}), we vary tropopause temperature to vary $V_p$. This method has an advantage to isolate structural changes of the TC owing to changes of $V_p$ without substantially affecting lower-tropospheric properties, such as $w_{cool}$. 
To further test the role of $C_d$ in modulating $M_{ew}$, two more sets of experiments CD and CDTTPP are designed (Table \ref{t1}). In CD, $C_d$ is varied while in CDTTPP, $C_d$ and $T_{tpp}$ are both varied in a manner that $V_p$ does not change (assuming the same air-sea enthalpy disequilibrium).
Parameter $C_k$ is further varied in CK (Table \ref{t1}) to test whether $M_{ew}$ is dominantly friction driven, as boundary layer thermodynamic property is expected to change in CK.
In the fifth set of experiments (FCOR, Table \ref{t1}), we vary Coriolis parameter $f$. Note CD, CDTTPP and CK will only be used for parameterizing $(M_{ew}/\rho_w)_{eq}$ (Sec. 2d), with their equilibrium periods all set to 40-50 days.

TCs are simulated using axisymmetric configuration of CM1 and the base state of the atmosphere is generated by three-dimensional simulations radiative-convective equilibrium without background rotation, same as Wang et al. (2022).  In FCOR and CD, base state of atmosphere are all the same as the case with $T_{tpp}=200$ K in TTPP. TCs are simulated for 50 days all simulations except 150 days for $f=1.25\times10^{-5}$ s$^{-1}$ and 100 days for $f=2.5\times10^{-5}$ s$^{-1}$ as the TCs with low $f$ takes longer to reach size equilibrium. Initial vortex for all experiments is the same and defined as in Rotunno and Emanuel (1987\nocite{Rotunno_Emanuel_1987}). Initial vortex maximum wind is about 13 m/s at a radius of about 100 km, see also Supplemental Material.

\begin{table}[t]
\caption{Parameters in experiments.}\label{t1}
\begin{center}
\begin{tabular}{ccccccccc}
\topline
 TTPP & CD & \multicolumn{2}{c}{CDTTPP} & CK & FCOR \\
\midline
$T_{tpp}$ (K) & $C_d$ & $T_{tpp}$ (K) & $C_d$ & $C_k$ & $f$ ($10^{-5}$ s$^{-1}$)\\
 241 & 0.0031 & 241 & 0.0007 & 0.0007 & - \\
 227 & 0.0023 & 227 & 0.0010 & 0.0010 & 1.25 \\
 214 & 0.0019 & 214 & 0.0012 & 0.0012 & 2.5 \\
 200 & 0.0015 & 200 & 0.0015 & 0.0015 & 5 \\
 187 & 0.0012 & 187 & 0.0018 & 0.0018 & 10 \\
 174 & 0.0010 & 174 & 0.0022 & 0.0022 & 20 \\
 163 & 0.0009 & 163 & 0.0025 & 0.0025 & - \\
\botline
\end{tabular}
\end{center}
\end{table}

Eyewall upward mass flux are approximated by the inflow under the eyewall:
\begin{equation}
    M_{ew} (t)=-2\pi r_{ew}\int_0^h\rho_d udz
\end{equation}
where $r_{ew}$ is some radius not far from the eyewall (here chosen as 2 times $r_{umin}$, the radius of minimum radial velocity in the boundary layer), $u$ the radial velocity, and $h$ the height of inflow layer, taken as the height where radial velocity $u$ is greater than 0.1 times the minimum $u$ at $r_{ew}$ ($=2r_{umin}$) following Zhang et al. (2011\nocite{Zhang_etal_2011}). 
$M_{ew}$ is processed by a 120-h running average. And $\rho_{w}$ is 
calculated such that $-2\pi r_{ew}\bar{h}\rho_wu_{avg}=M_{ew}$ in Eq. (C1), where $u_{avg}$ is the vertical mean radial velocity of the inflow after 120-h running average and where $\bar{h}$ is $h$ after 120-h running average.

To evaluate Eqs. (\ref{eq:kinematic}, \ref{eq:vten2}, \ref{eq:drtdt_0}), a spatial and temporal average is applied to remove noise in CM1 ouputs:
\begin{equation}\label{eq:rt_local_avg}
    \overline{[X]}= \frac{1}{z_2-z_1}\int_{z_1}^{z_2}[\frac{2}{(r_t+\Delta r)^2-(r_t-\Delta r)^2}\int_{r_t-\Delta r}^{r_t+\Delta r}(\frac{1}{P}\int_{t-P/2}^{t+P/2}Xdt) rdr]dz
\end{equation}
This average $\overline{[\cdot]}$ will apply to each term of Eq. (\ref{eq:vten2}): $\partial v/\partial t$ in Figs. \ref{fig:TTPP_size_comp}d, 
$u_t$ in Figs. \ref{fig:sens_Qlat_TTPP_56_34}, \ref{fig:sens_Qlat_FCOR};
$\Delta r=100$ km, $P=120$ h, $z_1=0$, $z_2=h_w=2.5$ km are set. 
Note one exception is that when applied to $\partial v/\partial r$ (Figs. \ref{fig:TTPP_size_comp}c, \ref{fig:FCOR_size_comp}d): we take the central difference using two radii $r_t-\Delta r$ and $r_t+\Delta r$ of the 120 h temporal running average of $v$, at a fixed height 1 km, for smoother results. 
As $\frac{\partial v}{\partial t}$ and $\frac{\partial v}{\partial r}$ are very noisy, they are further applied a 24-h running average after applying $\overline{[\cdot]}$ in Figs. \ref{fig:TTPP_size_comp}c-d and \ref{fig:FCOR_size_comp}d for clearer visualization. 

For calculation of $V_p$, air-sea enthalpy disequilibrium is taken as the radial average of the outer 900 km of the domain inside of the Rayleigh damping zone. Quantity $w_{cool}$ is calculated in the same manner as Wang et al. (2022). A typical value of $w_{cool}$ is 0.0027 m/s in the simulations.\footnote{Quantity $w_{cool}$ 
 was incorrectly calculated to be about half of the correct value in Wang et al. (2022).}

Quantity $\mathcal{Q}_{lat}$ from simulations (Figs. \ref{fig:TTPP_FCOR_Qlat_ov_2pirt}, \ref{fig:sens_Qlat_TTPP_56_34}, \ref{fig:sens_Qlat_FCOR}) is also needed: $\mathcal{Q}_{lat}$ is calculated as 120 h running average of  the mass integration (radially within $r_t$) of $c_p\Pi\Dot{\theta}_{pc}$ with $\theta$ the potential temperature, $\Pi$ the Exner function and $\Dot{\theta}_{pc}$ the potential temperature source due to phase changes (the tendency from microphysics section provided by CM1 subtracted by energy fallout term, see below).
The height of the volume for the integration  is defined as follows. First the location of maximum outflow velocity $u_{max}$ of 120 h running averaged fields is found. The height of volume is defined as the 1 km above the height where outflow velocity is less than 0.1$u_{max}$.

\clearpage

\appendix[D] 
\appendixtitle{Support for Eq. (\ref{eq:sdbud2}) and some parameter diagnosis and intepretation}

First, we provide a support of the dry-entropy balance equation Eq. (\ref{eq:sdbud2}). Recall that we define $s_d=c_p\ln\frac{\theta}{T_{trip}}$. Each term in Eq. (5) may be given as follows (see $\theta$ budget in Bryan and Rotunno 2009\nocite{Bryan_Rotunno_2009b} and CM1 governing equations in CM1 homepage), defining volume integration $\int_v=\int_0^{z_l}\int_0^{r_t}2\pi rdrdz$:

\begin{subequations}\label{eq:sdbud_terms}
\begin{equation}
    \frac{\partial \mathcal{S}}{\partial t}=\int_v\frac{\partial \rho_d s_d}{\partial t},
\end{equation}
\begin{equation}
\begin{split}
    \frac{\mathcal{Q}_{lat}}{T_{e,lat}}&=\int_v \{\rho_d[-\frac{c_v}{c_{vm}}(\frac{L_v}{T}\dot{q}_{gl}+\frac{L_f}{T}\dot{q}_{ls}+\frac{L_s}{T}\dot{q}_{gs})\\
    &\quad\quad\quad+(\frac{c_v}{c_{vm}}-\frac{R}{R_m})R_v(\dot{q}_{gl}+\dot{q}_{gs})]\}
\end{split}
\end{equation}
\begin{equation}
    -\frac{\mathcal{Q}_{rad}}{T_{e,rad}}=\int_v (\rho_d\frac{c_p}{\theta}\dot{\theta}_{rad}),
\end{equation}
\begin{equation}
    \dot{\mathcal{S}}_{res}=\int_v -\rho_d(\frac{c_v}{c_{vm}}R_m-R)\nabla \boldsymbol{u}+\int_v -\frac{c_v}{c_{vm}}\frac{\nabla\boldsymbol{J}}{T}+\int_v\frac{c_v}{c_{vm}}\frac{1}{T} (W_T+\epsilon),
\end{equation}
\begin{equation}
    \mathcal{F}_r=\int_0^{z_l} -2\pi r_t(\rho_d u s_d)|_{r=r_t} dz,
\end{equation}
\begin{equation}
    \mathcal{F}_u=\int_0^{r_t} -2\pi r(\rho_d w s_d)|_{z=z_l}dr,
\end{equation}
\end{subequations}
where $c_v$ the specific heat of dry air at constant volume, $c_{vm}=c_v+c_{vv}q_v+c_lq_l+c_sq_s$ with $c_{vv}$, $c_l$, $c_s$ the specific heat of water vapor, liquid water, solid water at constant volume, respectively, $q_l$, $q_s$ the mixing ratios of liquid water and solid water respectively; $R$ the gas constant of dry air, $R_m=R+q_vR_v$ the gas constant of moist air, with $R_v$ the gas constant of water vapor; $L_v$, $L_f$, $L_s$ the latent heat of vaporization, freezing and sublimation, respectively; 
$\dot{q}_{gl}$ and $\dot{q}_{gs}$ the $q_v$ source from phase changes between gas and liquid water and gas and solid water, respectively; 
$\dot{q}_{ls}$ the $q_l$ source from phase changes between liquid and solid water; $\dot{\theta}_{rad}$ the $\theta$ source due to radiative cooling; vector $\boldsymbol{u}$ the velocity;
vector $\boldsymbol{J}$ the sensible heat flux per unit area; $W_T$ the heating/cooling rate per unit volume due to falling hydrometeors and $\epsilon$ the dissipative heating. 
In particular, $W_T=-c_{vv}\boldsymbol{d}_v\nabla T-c_{l}\boldsymbol{d}_l\nabla T-c_{s}\boldsymbol{d}_s\nabla T+\boldsymbol{g}\cdot(\boldsymbol{d}_v+\boldsymbol{d}_l+\boldsymbol{d}_s)$, where $\boldsymbol{d}_v$, $\boldsymbol{d}_l$, $\boldsymbol{d}_s$ are diffusion (fall out) fluxes of water vapor, liquid water and solid water per unit area, respectively (see also Appendix A of Romps 2008, Wang and Lin 2021\nocite{Wang_Lin_2021}), $\boldsymbol{g}$ the gravitational acceleration.

In practice, Eq. (\ref{eq:sdbud_terms}) in a CM1 simulation is obtained by its automatic output of $\theta$ budget. 
Specifically, $\frac{\mathcal{Q}_{lat}}{T_{e,lat}}$ is obtained from CM1 output $ptb\_mp$ ($\theta$ source from microphysics section) subtracted by the effect of $W_T$ using CM1 output of terminal fall speed of different hydrometeors. 
Note in particular, $ptb\_mp$ itself can be a close approximation of $\theta$ source due to phase changes, because a simple scale estimation gives that dry-entropy source due to $W_T$ (Eq. \ref{eq:sdbud_terms}d) is only a few percent of that due to phase changes (Eq. \ref{eq:sdbud_terms}b) assuming $\partial T/\partial z=-7$ K/km and mean condensation height (for vertical distance of falling) of 3 km (this estimation is supported by explicit diagnosis not shown).

\begin{figure}[t]
 \noindent\includegraphics[width=35pc,angle=0]{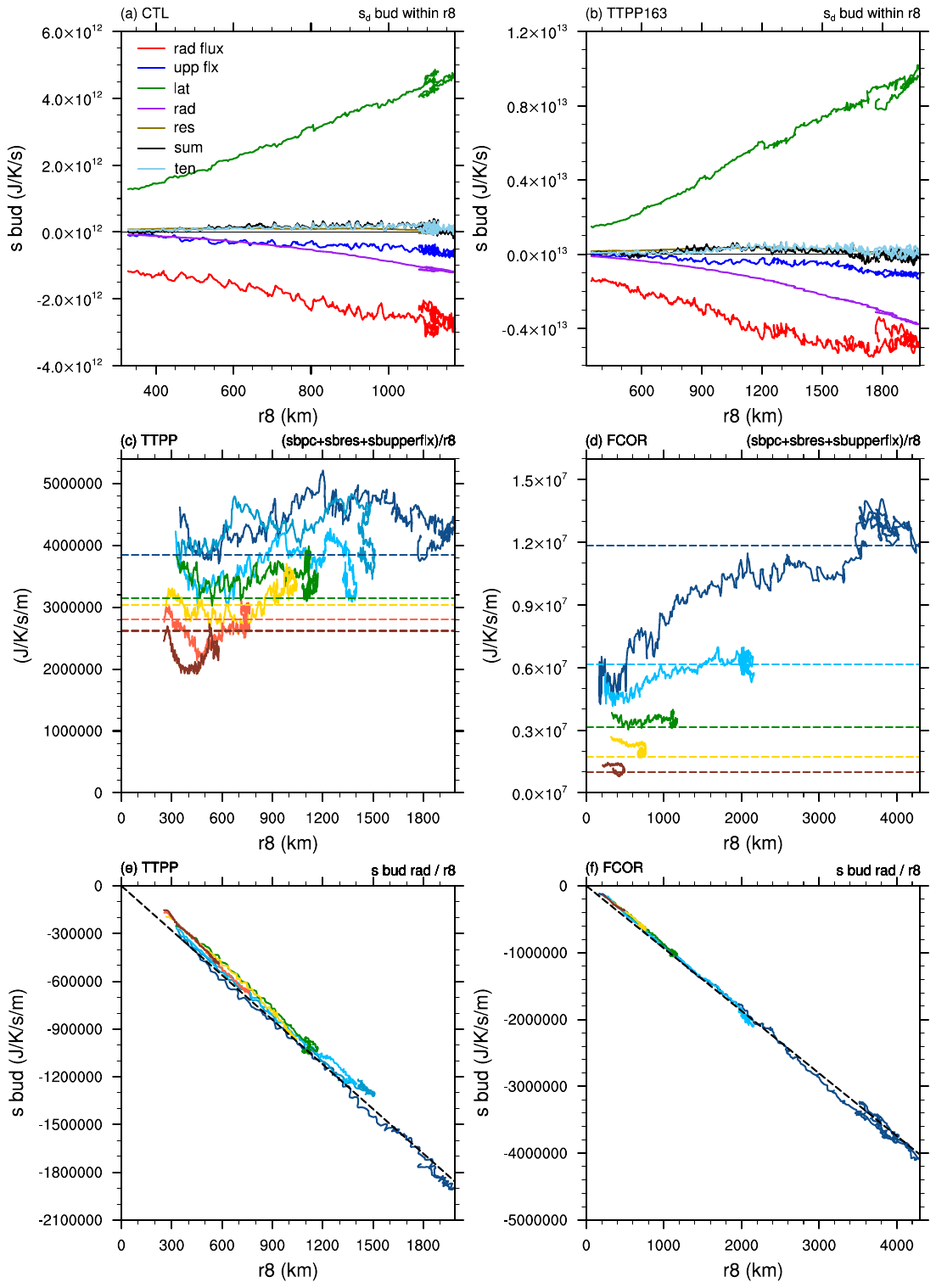}\\
 \caption{(a)-(b) terms in dry-entropy budget Eq. (\ref{eq:sdbud_terms}) (J/K/s) in CTL and $T_{tpp}=163$ K case in TTPP; legend shows (from top to bottom) terms $\mathcal{F}_r$, $\mathcal{F}_u$, $\frac{\mathcal{Q}_{lat}}{T_{e,lat}}$, $-\frac{\mathcal{Q}_{rad}}{T_{e,rad}}$, $\dot{\mathcal{S}}_{res}$, $\frac{\partial\mathcal{S}}{\partial t}$ as the sum these terms and directly calculated. (c)-(d) $(\frac{\mathcal{Q}_{lat}}{T_{e,lat}}+\dot{\mathcal{S}}_{res}+\mathcal{F}_u)/r_8$ (J/K/s/m) in TTPP and FCOR (solid lines), respectively, warmer color marking higher values of $T_{tpp}$ or $f$; dashed lines mark the expansion model predictions in Sec. 3b. (e)-(f) same as (c)-(d), but for $-\frac{\mathcal{Q}_{rad}}{T_{e,lat}}/r_8$ (J/K/s/m).
 }\label{fig:sdbud_r8_compare_TTPPFCOR}
\end{figure}

Fig. \ref{fig:sdbud_r8_compare_TTPPFCOR}a-b show support of the approximation given by Eq. (\ref{eq:sdbud2}) by representitive cases of CTL and $T_{tpp}=163$ K case in TTPP. It is evident that local tendency $\frac{\partial\mathcal{S}}{\partial t}$, dry-entropy source $\dot{\mathcal{S}}_{res}$, and vertical flux $\mathcal{F}_u$ are indeed negligible with dominant terms $\frac{\mathcal{Q}_{lat}}{T_{e,lat}}$, $\mathcal{F}_r$ and $-\frac{\mathcal{Q}_{rad}}{T_{e,rad}}$, which is important when TC is large. That a crucial assumption/approximation that $\frac{\mathcal{Q}_{lat}}{T_{e,lat}}$ scales with $r_t$ and $\frac{\mathcal{Q}_{rad}}{T_{e,rad}}$ scales with $r_t^2$ is more clearly (than in main text) supported by  Fig. \ref{fig:sdbud_r8_compare_TTPPFCOR}c-f.

\clearpage
\clearpage

\begin{figure}[t]
 \noindent\includegraphics[width=35pc,angle=0]{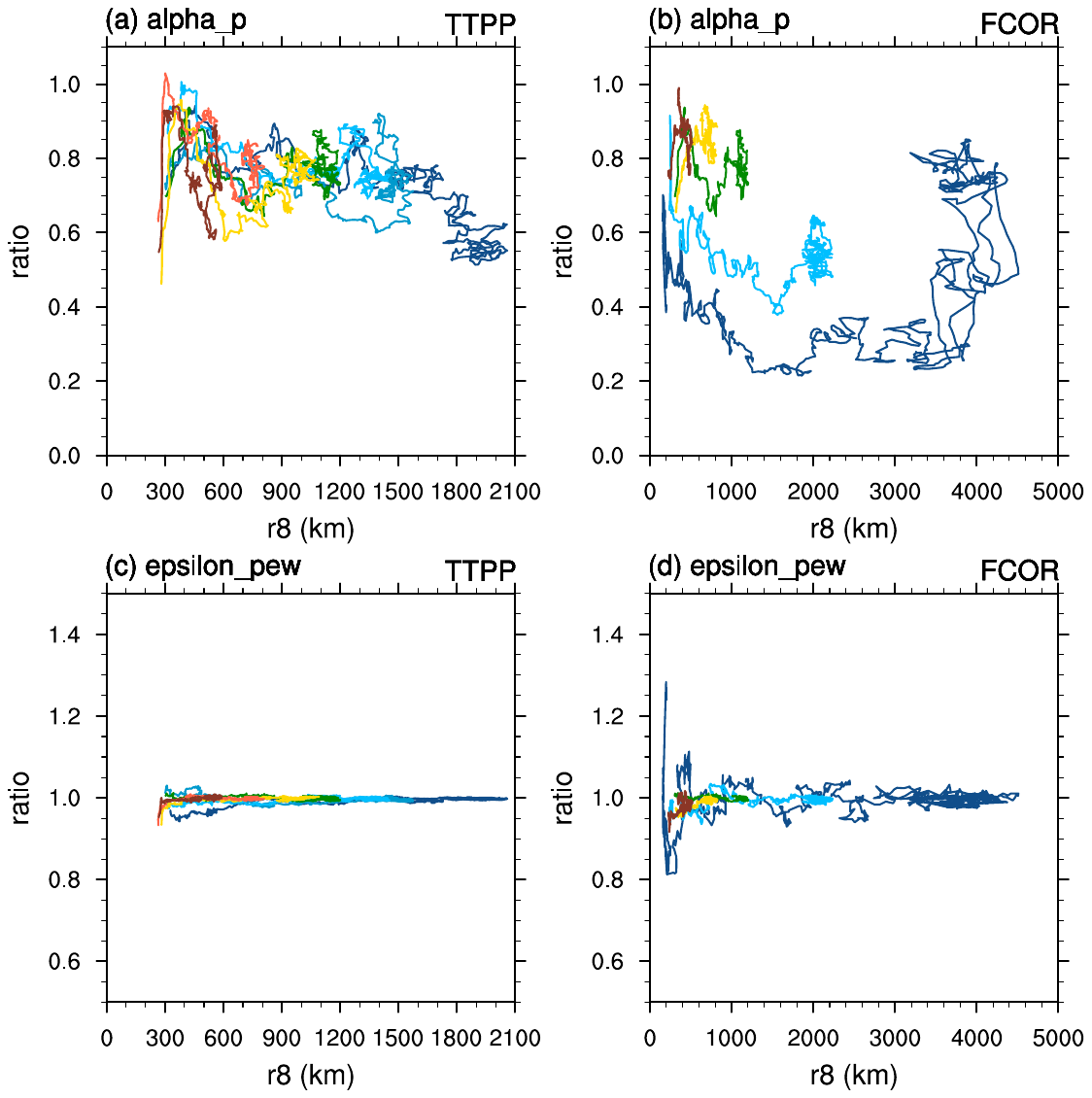}\\
 \caption{(a)-(b) Diagnosed $\alpha_p$ as a function of $r_8$ in TTPP and FCOR respectively. (c)-(d) same as (a)-(b) but for $\epsilon_{p,ew}$. See text for details.
 }\label{fig:alpha_p_epsilon_pew}
\end{figure}

In Sec. 3 we set $\alpha_p=0.8$, $\epsilon_{p,ew}=1$; here some support for this setting is shown in Fig. \ref{fig:alpha_p_epsilon_pew}. Consistent with the definition in Sec. 2a, $\alpha_p$ is calculated as the ratio between the latent heating within two times radius of maximum wind ($r_m$) to $Q_{lat}$. Two times of $r_m$ is to account for the slope of the eyewall, which could also include some inner rainband according to Wang (2009\nocite{Wang_2009}). It is seen that $\alpha_p$ in TTPP is approximately constant being about 0.8, suggesting the dominant contribution of latent heating in the eyewall to total latent heating within $r_8$. In FCOR, however, it is seen $\alpha_p$ can be substantially smaller when $f=2.5\times10^{-5}$ and $f=1.25\times10^{-5}$ s$^{-1}$, indicating that latent heating outside of the eyewall is important in driving expansion of these cases. It is also noted that when $f=2.5\times10^{-5}$ s$^{-1}$, $\alpha_p$ is still mainly above 0.6 when $r_8$ is smaller than 1000 km, a radius more relevant to TCs on earth.

Here for convenience $\epsilon_{p,ew}$ is diagnosed/estimated as the ratio of 120-h running averaged precipitation to the 120-h running averaged sum of surface water vapor and lateral water (vapor and hydrometeors) fluxes within (at) two times of $r_m$. Note $\epsilon_{p,ew}$ thus defined is similar to the large scale precipitation efficiency in Sui et al. (2005). 
It is seen in Fig. \ref{fig:alpha_p_epsilon_pew}c-d that $\epsilon_{p,ew}$ is about 1 during whole expansion stage. This indicates negligible local accumulation of water in the atmosphere and nearly all water vapor input to the eyewall changes phase.

\clearpage
\clearpage

\begin{figure}[t]

\noindent\includegraphics[width=20pc,angle=0]{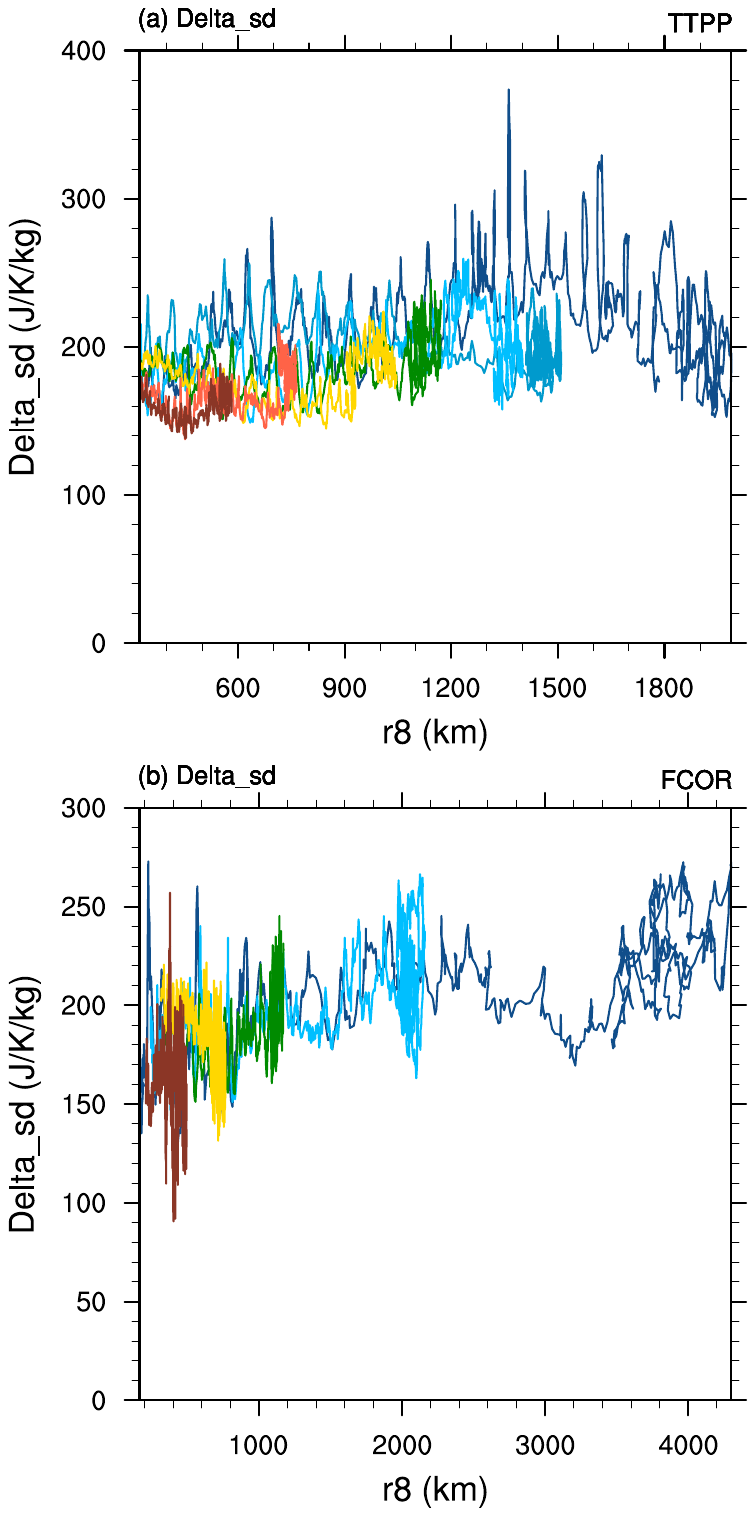}\\
 \caption{Diagnosed $\Delta s_d$ (J/K/kg) in (a) TTPP and (b) FCOR using Eq. (\ref{eq:delta_sd}). Colors have the same meaning as Fig. D1. Note in calculation the numerator and denominator of Eq. (\ref{eq:delta_sd}) are first processed by a 120-h running average. 
 }\label{fig:delta_sd_TTPP_FCOR}
\end{figure}

We also show $\Delta s_d$ diagnosed by Eq. (\ref{eq:delta_sd}) in simulations in Fig. \ref{fig:delta_sd_TTPP_FCOR}. It is seen that diagnosed $\Delta s_d$ are very close in different experiments of TTPP and FCOR and are also approximately constant with time. The value is also not far from the 187.1 J/K/kg estimation in the ideal base environment in Sec. 3. This supports our assumption of a constant $\Delta s_d$ and indirectly supports the interpretation of $\Delta s_d$. Our expectation that $\Delta s_d$ is mainly driven by sea surface temperature needs explicit simulation to verify. Note diagnosed $\Delta s_d$ slightly increases with decreasing $T_{tpp}$; this also means the faster expansion rate with lower $T_{tpp}$ is not caused by a $\Delta s_d$ sensitivity. 

The meaning of $\Delta s_d$ may be further understood in a more ideal picture where inflow at $r_t$ occur only from surface to height $z_i$ and outflow only occur from $z_i$ to $z_l$ and where $z_i\lesssim z_l$ so that outflow is well confined to tropopause level. 
To simplify math, we also assume buoyancy frequency $N$ is a constant so that $\frac{\partial s_d}{\partial z}=\frac{c_p}{g}N^2$ is a constant (not a bad approximation as seen in CTL, see Supplemental Material). We denote $s_d$ at surface to be $s_{d0}$ so that $s_{di}=s_{d0}+\frac{c_p}{g}N^2z_i$ is the $s_d$ at height $z_i$; we denote $s_{d,tpp}$ the $s_d$ at height $z_l$. Thus, $\mathcal{F}_r$ (Eq. \ref{eq:sdbud_terms}e) is written as:
\begin{equation}\label{eq:Fr2}
\mathcal{F}_r=\int_{0}^{z_i}-2\pi r_t(\rho_d us_d)dz+\int_{z_i}^{z_l}-2\pi r_t(\rho_d us_d)dz
\end{equation}
The first term on the RHS is written:
\begin{equation}
\begin{split}
&\quad\int_0^{z_i}-2\pi r_t(\rho_dus_d)dz\\
&=-2\pi r_t\int_0^{z_i}\rho_du(s_{d0}+\frac{c_p}{g}N^2z)dz\\
&=s_{d0}(-2\pi r_t\int_0^{z_i}\rho_dudz)-2\pi r_t\frac{c_p}{g}N^2\int_0^{z_i}\rho_duzdz\\
&=s_{d0}\psi_i+\frac{c_p}{g}N^2\int_0^{z_i}\frac{\partial \psi}{\partial z}zdz\\
&=s_{d0}\psi_i+\frac{c_p}{g}N^2[(\psi z)\big|_0^{z_i}-\int_0^{z_i}\psi dz]\\
&=s_{d0}\psi_i+\frac{c_p}{g}N^2\psi_iz_i-\frac{c_p}{g}N^2\int_0^{z_i}\psi dz\\
&=s_{di}\psi_i-\frac{c_p}{g}N^2\int_0^{z_i}\psi dz
\end{split}
\end{equation}
where $\psi$ is mass streamfunction.
The second term on the RHS of Eq. (\ref{eq:Fr2}) is written:
\begin{equation}
\int_{z_i}^{z_l}-2\pi r_t(\rho_dus_d)dz=\int_{z_i}^{z_l}\frac{\partial\psi}{\partial z}s_ddz\approx s_{di}(-\psi_i)
\end{equation}
Thus, we have $\mathcal{F}_r=-\frac{c_p}{g}N^2\int_0^{z_i}\psi dz=-\frac{c_p}{g}N^2\bar\psi z_i$. Note also that $2\pi r_t\rho_iu_th_w=-\psi_{h_w}$. Then $\Delta s_d$ (Eq. \ref{eq:delta_sd}) is:
\begin{equation}
\begin{split}
\Delta s_d&=\mathcal{F}_r/(-\psi_{h_w})\\
&=\frac{c_p}{g}N^2z_i\frac{\bar\psi}{\psi_{h_w}}\\
&\approx(s_{d,tpp}-s_{d0})\frac{\bar\psi}{\psi_{h_w}}
\end{split}
\end{equation}
Thus, we see that $\Delta s_d$ will represent the difference of $s_d$ between tropopause and surface if $\psi_{h_w}$ is close to the vertical mean of $\psi$ in the whole inflow layer. A structure with inflow confined near surface satisfies this condition; but other vertical profiles of inflow can also be valid.

\bibliographystyle{ametsocV6}
\bibliography{refs_CHAVAS_add}

\end{document}